\documentclass[useAMS,usenatbib]{mn2e}

\usepackage{color}
\usepackage{soul}  
\usepackage{mathrsfs} 
\usepackage{amssymb}
\usepackage{ctable} 
\usepackage{hyperref}
\usepackage{epsfig}
\usepackage{epstopdf}

\renewcommand{\topmargin}{-1.7cm}
\title[Cosmic homogeneity in the WiggleZ survey]{The WiggleZ Dark Energy Survey: the transition to large-scale cosmic homogeneity}

\author[M. I. Scrimgeour et al.]
{\parbox{\textwidth}{Morag I. Scrimgeour$^{1,2}$\thanks{E-mail: 
\texttt{morag.scrimgeour@icrar.org}},
Tamara Davis$^{3}$,  Chris Blake$^{4}$, J. Berian James$^{5,6}$, \\ Gregory B. Poole$^4$, Lister Staveley-Smith$^{1,2}$, Sarah Brough$^7$,  Matthew Colless$^{7}$,  Carlos Contreras$^4$, Warrick Couch$^{4}$, Scott Croom$^{8}$, 
Darren Croton$^4$, \\ Michael J. Drinkwater$^3$, Karl Forster$^9$, David Gilbank$^{10}$, Mike Gladders$^{11}$, \\ Karl Glazebrook$^{4}$, Ben Jelliffe$^8$, Russell J. Jurek$^{12}$, I-hui Li$^4$, Barry Madore$^{13}$, \\ D. Christopher Martin$^9$,  Kevin Pimbblet$^{14}$, Michael Pracy$^{8}$, Rob Sharp$^{7,15}$, \\ Emily Wisnioski$^4$,  David Woods$^{16}$, Ted K. Wyder$^9$ and H.K.C. Yee$^{17}$}\vspace{0.4cm}\\
\parbox{\textwidth}{
$^1$ International Centre for Radio Astronomy Research, M468, University of Western Australia, 35 Stirling Hwy, Crawley, WA 6009, Australia \\
$^2$ ARC Centre of Excellence for All-sky Astrophysics (CAASTRO) \\
$^3$ School of Mathematics and Physics, University of Queensland, QLD 4072, Australia\\
$^4$ Centre for Astrophysics \& Supercomputing, Swinburne University of Technology, P.O. Box 218, Hawthorn, VIC 3122, Australia \\
$^5$ Dark Cosmology Centre, Juliane Maries Vej 30, 2100 Copenhagen $\O$, Denmark \\
$^6$ Astronomy Department, 601 Campbell Hall, Berkeley CA 94720, USA \\
$^7$ Australian Astronomical Observatory, P.O. Box 296, Epping, NSW 1710, Australia \\
$^8$ Sydney Institute for Astronomy, School of Physics, University of Sydney, NSW 2006, Australia \\
$^9$ California Institute of Technology, MC 278-17, 1200 East California Boulevard, Pasadena, CA 91125, United States \\
$^{10}$ Astrophysics and Gravitation Group, Department of Physics and Astronomy, University of Waterloo, Waterloo, ON N2L 3G1, Canada \\
$^{11}$ Department of Astronomy and Astrophysics, University of Chicago, 5640 South Ellis Avenue, Chicago, IL 60637, United States \\
$^{12}$ Australia Telescope National Facility, CSIRO, Epping, NSW 1710, Australia \\
$^{13}$ Observatories of the Carnegie Institute of Washington, 813 Santa Barbara St., Pasadena, CA 91101, United States \\
$^{14}$ School of Physics, Monash University, Clayton, VIC 3800, Australia \\
$^{15}$ Research School of Astronomy \& Astrophysics, Australian National University, Weston Creek, ACT 2600, Australia\\
$^{16}$ Department of Physics \& Astronomy, University of British Columbia, 6224 Agricultural Road, Vancouver, BC V6T 1Z1, Canada\\
$^{17}$ Department of Astronomy and Astrophysics, University of Toronto, 50 S. George Street, Toronto, ON M5S 3H4, Canada}}

\date{Released 2011 Xxxxx XX}

\pagerange{\pageref{firstpage}--\pageref{lastpage}} \pubyear{2002}

\def\LaTeX{L\kern-.36em\raise.3ex\hbox{a}\kern-.15em
    T\kern-.1667em\lower.7ex\hbox{E}\kern-.125emX}

\begin{document}

\label{firstpage}

\maketitle

\begin{abstract}
We have made the largest-volume measurement to date of the transition to large-scale homogeneity in the distribution of galaxies. We use the WiggleZ survey, a spectroscopic survey of over 200\,000 blue galaxies in a cosmic volume of $\sim1$ $h^{-3}$Gpc$^3$. A new method of defining the `homogeneity scale' is presented, which is more robust than methods previously used in the literature, and which can be easily compared between different surveys.  Due to the large cosmic depth of WiggleZ (up to $z=1$) we are able to make the first measurement of the transition to homogeneity over a range of cosmic epochs. The mean number of galaxies $N(<r)$ in spheres of comoving radius $r$ is proportional to $r^3$ within 1 per cent, or equivalently the fractal dimension of the sample is within 1 per cent of $D_2=3$, at radii larger than $71 \pm8$ $h^{-1}$Mpc at $z\sim0.2$, $70 \pm 5$ $h^{-1}$Mpc at $z\sim0.4$, $81 \pm5$ $h^{-1}$Mpc at $z\sim0.6$, and $75 \pm4$ $h^{-1}$Mpc at $z\sim0.8$.  We demonstrate the robustness of our results against selection function effects, using a $\Lambda$CDM $N$-body simulation and a suite of inhomogeneous fractal distributions. The results are in excellent agreement with both the $\Lambda$CDM $N$-body simulation and an analytical $\Lambda$CDM prediction. We can exclude a fractal distribution with fractal dimension below $D_2=2.97$ on scales from $\sim80$ $h^{-1}$Mpc up to the largest scales probed by our measurement, $\sim$300 $h^{-1}$Mpc, at 99.99 per cent confidence.

\end{abstract}


\section{Introduction}
\label{intro}

One of the main assumptions of the standard theory of cosmology, $\Lambda$CDM (based on cold dark matter and a cosmological constant), is that the Universe is homogeneous and isotropic on large scales, and hence can be described by the Friedmann-Robertson-Walker (FRW) metric. `Homogeneous' means that its statistical properties (such as density) are translationally invariant; `isotropic' means it should be rotationally invariant. The Universe clearly deviates from this on small scales, where galaxies are clustered, but on large enough scales ($\gtrsim100h^{-1}$Mpc in $\Lambda$CDM), the distribution of matter is assumed to be `statistically homogeneous' --  i.e., the small-scale inhomogeneities can be considered as perturbations, which have a statistical distribution that is independent of position. However, this is merely an assumption, and it is important for it to be accurately verified by observation. Over the last decade there has been a debate in the literature as to whether the Universe really is homogeneous, or whether it has a fractal-like structure extending to large scales. It is important to resolve this contention if we are to be justified in assuming the FRW metric.

In fact, although $\Lambda$CDM is based on the assumption of large-scale homogeneity and an FRW metric, inflation (which $\Lambda$CDM incorporates) actually predicts a certain level of density fluctuations on all scales. Inflation predicts that the primordial density power spectrum was close to scale-invariant. In the standard model, the scalar index $n_s$, which quantifies the scale-dependence of the primordial power spectrum, is close to 0.96 \citep{baumann2009}, while a scale-invariant power spectrum has $n_s=1$. In this case, these density fluctuations induce fluctuations in the metric, $\delta \Phi$, which are virtually independent of scale, and are on the order of $\delta \Phi / c^2 \sim 10^{-5}$ \citep{peacock1999}. Since these perturbations are small, the FRW metric is still valid, but it means that we expect the Universe to have a gradual approach to large-scale homogeneity rather than a sudden transition.

The most important implication of inhomogeneity is the so-called `averaging problem' in General Relativity (GR). This arises when we measure `average' quantities (such as the correlation function and power spectrum, and parameters such as the Hubble constant) over a spatial volume. In doing so we assume the volume is homogeneous and smooth, when it may not be. Since the Einstein equations are nonlinear, density fluctuations can affect the evolution of the average properties of the volume -- this is known as the `backreaction mechanism' (e.g. \cite{buchert2000}, \cite{ellis2005}, \cite{li2007}; see \cite{rasanen2011} for a summary). If we observe a quantity within such a volume, we need to take averaging into account to compare it with theory.  It is therefore important to know how much inhomogeneity is present, in order to obtain meaningful results from averaged measurements (and so most, if not all, cosmological measurements). 

Backreaction has also been proposed as an explanation of dark energy, which is believed to be a negative-pressure component of the Universe that drives the accelerated expansion. Some authors have suggested that instead of introducing exotic new forms of dark energy, or modifications to GR, we should revisit the fundamental assumptions of the $\Lambda$CDM model, such as homogeneity. If we assume that GR holds, but take inhomogeneities into account, it can be shown that backreaction can cause a global cosmic acceleration, without any additional dark energy component \citep[see e.g.][]{schwarz2002,kolb2005,rasanen2006,wiltshire2007a,buchert2008,rasanen2011}. This effect appears to be too small to explain the observed acceleration, but highlights the importance of understanding the amount of inhomogeneity in the Universe.

Another, related, consequence of inhomogeneity is that it can affect the path travelled by light rays, and the calibration of clocks and rods of observers. It can therefore affect distance measurements, such as redshifts and luminosity distances \citep{wiltshire2009,meures2012}. This has also been proposed as a possible explanation of the observed cosmic acceleration, although the effect appears to be on the order of only a few percent at $z\sim1$ \citep{brouzakis2007}.

Homogeneity is required by several important statistical probes of cosmology, such as the galaxy power spectrum and $n$-point correlation functions, in order for them to be meaningful. Applying these to a galaxy sample below the scale of homogeneity would be problematic, since if a distribution has no transition to homogeneity, it does not have a defined mean density, which is required to calculate and interpret these statistics. It is also not possible to model its cosmic variance, so the error in the measurements would be ill-defined, making it impossible to relate these statistics to a theoretical model.  It is therefore important to quantify the scale on which the Universe becomes close enough to homogeneous to justify their use. 

Large-scale homogeneity is already well supported by a number of different observations. In particular, the high degree of isotropy of the CMB \citep{fixsen1996} gives very strong support for large-scale homogeneity in the early Universe, at redshift $z \sim 1100$. The isotropy of the CMB also indicates the Universe has remained homogeneous, since there are no significant Integrated Sachs-Wolfe (ISW) effects distorting our view of the isotropic CMB \citep{wu1999}. Other high-redshift evidence for homogeneity includes the isotropy of the X-Ray Background (XRB)  \citep{peebles1993,scharf2000}, believed to be emitted by high-redshift sources, and the isotropy of radio sources at $z \sim 1$ \citep{blake2002}.

However, these measurements of high-redshift isotropy do not necessarily imply homogeneity of the present Universe. If every point in the Universe is isotropic, then this implies the Universe is homogeneous; so if we accept the Copernican principle that our location is non-special, then the observed isotropy should imply homogeneity \citep{peacock1999}.\footnote{We note that the Copernican principle is not incompatible with an inhomogeneous Universe. It assumes only that our location is non-special, not that every location is the same \citep{joyce2000,clifton2008,syloslabini2009}.} However most of these measurements (except the ISW effect) only tell us about the high-redshift Universe. We know that it has evolved to a clustered distribution since then, and it is possible that it could also have become anisotropic. It is also possible for the matter distribution to be homogeneous while the galaxy distribution is not, since the galaxy distribution is biased relative to the matter field \citep{kaiser1984} -- although since galaxy bias is known to be linear on large scales \citep{coles1993,scherrer1998}, this seems unlikely. The ISW effect \citep{sachs1967} gives information about the low-redshift Universe, since it mostly probes the dark energy dominated era, $z\lesssim1$ \citep{afshordi2004}, but it is an integral over the line-of-sight and so does not give full 3D information. 

Galaxy surveys are the only 3D probe of homogeneity in the nearby Universe, and a number of homogeneity analyses have been carried out with different surveys, with seemingly conflicting results. Most statistical methods used to measure homogeneity have been based on the simple `counts-in-spheres' measurement, that is, the number of galaxies $N(<r)$ in spheres of radius $r$ centred on galaxies, averaged over a large number of such spheres. This quantity scales in proportion to the volume ($r^3$) for a homogeneous distribution, and homogeneity is said to be reached at the scale above which this holds. \cite{hogg2005} applied this to the Sloan Digital Sky Survey (SDSS) luminous red galaxy (LRG)
 sample at $z\sim0.3$, and found the data became consistent with homogeneity at $\sim70 h^{-1}$ Mpc for this galaxy population (using a different method of determining the homogeneity scale than we do).

This measurement can also be extended to a fractal analysis. Fractal dimensions can be used to quantify clustering; they quantify the scaling of different moments of galaxy counts in spheres, which in turn are related to the $n$-point correlation functions. The most commonly used is the correlation dimension $D_2(r)$, which quantifies the scaling of the 2-point correlation function, and is based on the counts-in-spheres, which scale as $N(<r) \sim r^{D_2}$. One can also consider the more general dimensions $D_q$, where $q$ are different moments of the counts-in-spheres. Using fractal analyses, some researchers have found a transition from $D_2<3$ to $D_2=3$, at around 70 -- 150 $h^{-1}$ Mpc \citep{martinez1994,guzzo1997,martinez1998,scaramella1998,amendola1999,pan2000,kurokawa2001,yadav2005,sarkar2009}, whereas other authors have found no such transition \citep{coleman1992,pietronero1997,syloslabini1998,joyce1999,syloslabini2009, syloslabini2011}. However, many of  the galaxy redshift surveys used in the above-mentioned works are too shallow, sparse, or have survey geometries too complicated, to give conclusive results.

In this work, we use the WiggleZ Dark Energy Survey \citep{drinkwater2010} to make a new measurement of the counts-in-spheres and correlation dimension, to test for the transition to homogeneity. WiggleZ provides a larger volume than previous surveys, making it ideal for a homogeneity measurement, and it covers a higher redshift range, allowing us to also investigate how homogeneity changes with cosmic epoch. It is not volume-limited, but we show that this does not significantly affect our measurement. The transition to homogeneity can be used as a test of a particular cosmological model, since we would expect it to differ for different cosmologies. In this work we test a $\Lambda$CDM model with best-fitting parameters from the Wilkinson Microwave Anisotropy Probe (WMAP) data \citep{komatsu2011}, which we refer to as $\Lambda$CDM+WMAP. We demonstrate the robustness of our measurement against systematic effects of the survey geometry and selection function, by repeating our analysis on both the GiggleZ $N$-body simulation and on a suite of inhomogeneous, fractal distributions.

Before we make any meaningful test of homogeneity, however, it is crucial to properly define what we mean by the so-called `scale' of homogeneity. Since there is only a gradual approach to homogeneity, such a definition may be arbitrary. In the past, authors have defined the `scale of homogeneity' as the scale where the data becomes consistent with homogeneity within 1$\sigma$ \citep[e.g.][]{hogg2005,bagla2008,yadav2010}. However, this method has several disadvantages. It depends on the size of the error bars on the data, and hence on the survey size. A larger survey should have smaller error bars, and so will automatically measure a larger scale of homogeneity. It also depends on the bin spacing, and is susceptible to noise between data points. We therefore introduce a different, and more robust, method for determining homogeneity: we fit a smooth, model-independent polynomial curve to all the data points, and find where this intercepts chosen values close to homogeneity.

Certain parts of our analysis require the assumption of a cosmological model and, implicitly, homogeneity (i.e. for converting WiggleZ redshifts to distances, correcting for the selection function, calculating the uncertainties using lognormal realisations, and finding the best-fitting bias). In these cases, we use an input $\Lambda$CDM cosmology with $h=0.71$, ${\Omega_m=0.27}$, $\Omega_\Lambda=0.73$, $\Omega_b=0.04482$, $\sigma_8=0.8$ and $n_s=0.96$. Here, the Hubble constant is ${H_0 = 100 h \textrm{ km s}^{-1} \textrm{ Mpc}^{-1}}$, $\Omega_m$ is the mass density, $\Omega_\Lambda$ is the dark energy density, $\Omega_b$ is the baryon density, $\sigma_8$ is the root mean square mass variation within spheres of 8$h^{-1}$Mpc radius, and $n_s$ is the spectral index of the primordial power spectrum. This is the same fiducial cosmology used by \cite{blake2011a}, and we use this for consistency.  We discuss the implications of assuming a  $\Lambda$CDM model on the results of our homogeneity measurement in Section \ref{discussion}.

This paper is organised as follows.  In Section \ref{survey} we describe the WiggleZ survey and our dataset. In Section \ref{methodology} we describe our methodology. We also explain our definition of the `scale of homogeneity' and present a new model-independent method for measuring this from data. In Section \ref{analytic} we describe our analytic $\Lambda$CDM+WMAP model. We present our results in Section \ref{results}. We test the robustness of our method using fractal distributions and a $\Lambda$CDM $N$-body simulation in Section \ref{robustness}. We discuss our results in Section \ref{discussion} and conclude in Section \ref{conclusion}.

\section{The WiggleZ Survey}
\label{survey}

The WiggleZ Dark Energy Survey \citep{drinkwater2010} is a large-scale spectroscopic galaxy redshift survey conducted at the 3.9m Anglo-Australian Telescope, and was completed in January 2011. It maps a cosmic volume of $\sim 1$ Gpc$^3$ up to redshift ${z=1}$, and has obtained $239\,000$ redshifts for UV-selected emission-line galaxies with a median redshift of $z_{\rm med}=0.6$. Of these, 179\,599 are in regions contiguous enough to be used for our analysis. It covers $\sim1000$ deg$^2$ of equatorial sky in 7 regions, shown in \cite{drinkwater2010} (their Fig.1).

The observing strategy and galaxy selection criteria of the WiggleZ survey are described in \cite{blake2009} and \cite{drinkwater2010}. The selection function we use is described in \cite{blake2010}. The targets were selected from Galaxy Evolution Explorer satellite (GALEX) observations matched with ground-based optical measurements, and magnitude and colour cuts were applied to preferentially select blue, extremely luminous high-redshift star-forming galaxies with bright emission lines. 

The WiggleZ Survey offers several advantages for a new study of the scale of homogeneity. Its very large volume allows homogeneity to be probed on scales that have not previously been possible, and at a higher redshift; it probes a volume at ${z>0.5}$ comparable to the SDSS LRG catalogue at ${z<0.5}$. This allows us to make the first measurement of the change in the homogeneity scale over a range of cosmic epochs.  We divide our sample into four redshift slices, $0.1<z<0.3$, $0.3<z<0.5$, $0.5<z<0.7$ and $0.7<z<0.9$. The sizes of the WiggleZ regions in each redshift slice are listed in Table \ref{regionsizes}; they sample scales well above the expected scale of homogeneity.  The numbers of galaxies in each redshift slice are listed in Table \ref{regionnumbers}.  

We also benefit from having 7 regions distributed across the equatorial sky, which reduces the effect of cosmic variance. In addition, WiggleZ probes blue galaxies, whereas SDSS (which has obtained the largest-scale measurements of homogeneity to date) used Luminous Red Galaxies, and so it can also constrain any systematic effects introduced by the choice of tracer galaxy population \citep{drinkwater2010}.  Since blue galaxies are less biased, they are also more representative of the underlying matter distribution.

\begin{table*}
 \caption{Comoving dimensions of WiggleZ regions in each redshift slice, in units of [$h^{-1}$ Mpc]. The dimensions correspond to the line-of-sight, RA, and Dec directions, respectively. }
 \label{regionsizes}
 \begin{tabular}{@{}lcccccc}
  \hline
Region & $0.1<z<0.3$ &$0.3<z<0.5$ &  $0.5<z<0.7$&$0.7<z<0.9$  \\
  \hline
 00-hr & $551 \times146 \times222 $ & $505\times232\times353$ & $459\times309\times471$  & $417\times378\times575$   \\
01-hr & $ 549 \times 127 \times 132 $ & $ 499 \times  202 \times 209 $ & $450 \times 269 \times 279 $ & $ 405\times329 \times 341$   \\
 03-hr & $ 549\times133 \times 129$ & $ 499\times 211 \times206 $ & $ 450\times281 \times 274$ & $ 405\times343 \times 335$   \\
09-hr & $551 \times 221\times 133$ & $504 \times351 \times212 $ & $ 458\times467\times283 $ & $416 \times 571\times346 $   \\
11-hr & $553 \times280 \times145 $ & $509 \times445 \times231 $ & $ 466\times592 \times308 $ & $426 \times724 \times 376$   \\
15-hr & $ 553\times295 \times 150$ & $510 \times468 \times 238$ & $468 \times623 \times317 $ & $ 429\times762 \times 387$   \\
22-hr & $ 550\times142 \times 143$ & $ 500\times225 \times 228$ & $452 \times300 \times 303$ & $408 \times367 \times371 $   \\
  \hline
 \end{tabular}
 \end{table*}

 \begin{table}
  \begin{center}
 \caption{Number of WiggleZ galaxies in each redshift slice. }
 \label{regionnumbers}
 \begin{tabular}{@{}ccccccc}
  \hline
Redshift & Number of galaxies  \\
  \hline
  $0.1<z<0.3$ & 25\,187 \\
$0.3<z<0.5$ & 45\,698 \\
 $0.5<z<0.7$ & 70\,191\\
$0.7<z<0.9$ & 38\,523  \\
  \hline
 \end{tabular}
 \end{center} 
 \end{table}

There are, however, several aspects of the survey that could potentially be detrimental to a homogeneity measurement. WiggleZ has a complex window function, with a complicated edge geometry including holes in the angular coverage, and the spectroscopic completeness varies across the sky. In addition, the population properties of the galaxies are known to vary with redshift, due to the effects of Malmquist bias (since WiggleZ is a flux-limited survey), downsizing \citep[the observed fact that the size of the most actively star-forming galaxies decreases with time, ][]{ cowie1996,glazebrook2004}, and the colour and  magnitude selection cuts \citep{blake2010}. This means that WiggleZ preferentially selects larger-mass, higher-luminosity galaxies at higher redshift. A consequence of this is that it is not possible to define volume-limited subsamples of WiggleZ. However, we can correct for these effects by using random catalogues (Section \ref{nr}), which account for the survey selection function \citep{blake2010}. We also divide the survey into four redshift-slices, reducing the amount of galaxy population evolution in any region. In addition, we show that our results are not biased by the assumption of homogeneity in the selection function corrections, using an $N$-body simulation and a suite of inhomogeneous fractal distributions, described in Section \ref{robustness}. We are therefore confident that our result is not distorted by any features of the survey. 

We convert the redshifts of the WiggleZ galaxies to comoving distances $d_c$, using
\begin{equation}
d_c(z) = {c \over H_0}  \int ^{z} _{0} {dz' \over E(z')},
\end{equation}
where
\begin{equation}
\label{Ez}
E(z) = {H(z) \over H_0} =  [\Omega_{m,0} (1+z)^{3} + \Omega_{\Lambda,0}]^{1/2},
\end{equation}
and we use the fiducial $\Lambda$CDM parameter values listed in Section 1. To do this, we assume the FRW metric and $\Lambda$CDM. This is necessary for any homogeneity measurement, since we must always assume a metric in order to interpret redshifts. Therefore in the strictest sense this can only be used as a consistency test of $\Lambda$CDM. However, if we find the trend towards homogeneity matches the trend predicted by $\Lambda$CDM, then this is a strong consistency check for the model and one that an inhomogeneous distribution would find difficult to mimic. We discuss this further in Section \ref{discussion}.

\section{Methodology}
 
Here we describe our methodology for measuring the transition to homogeneity. We first calculate the mean counts-in-spheres $N(<r)$, then we find the fractal correlation dimension $D_2(r)$, from the slope of $N(r)$. Although they are closely related it is interesting to consider both, since the counts-in-spheres is the simplest measurement of homogeneity, whilst the correlation dimension provides direct information about the fractal properties of the distribution. We also describe our method of determining uncertainties, and our method of defining the `homogeneity scale' $R_H$.
\label{methodology}

\subsection{Scaled counts-in-spheres ${\mathcal{N}(<r)}$}
\label{nr}

The simplest test of homogeneity of a set of points is to find the average number $N(<r)$ of neighbouring points from any given point, up to a maximum distance $r$; if the distribution is homogeneous, then (for large enough $r$),
\begin{equation}
N(<r) \propto r^D,
\end{equation}
where $D$ is the ambient dimension (the number of dimensions of the space; for a homogeneous volume, $D=3$).

We find ${N(<r)}$ for spheres centred on each of the WiggleZ galaxies, and correct for incompleteness by dividing by the number expected for a homogeneous distribution with the same level of completeness. (We show that this does not bias our final results). This is done by finding the mean ${N(<r)}$ about the coordinate position of the WiggleZ galaxy from 100 random catalogues, each with the same number density, window function and redshift distribution of the WiggleZ survey. The method of generating the random catalogues is described in \cite{blake2010}. We then take the average over all the galaxies, to obtain the mean, scaled counts-in-spheres measurement ${\mathcal{N}(<r)}$:

\begin{equation}
\label{nr_eq}
\mathcal{N}(<r) = {1 \over G} \sum_{i=1}^G {N^i(<r) \over {1 \over R} \sum_{j=1}^R \rho_j N_{R}^{i,j}(<r)} ,
\end{equation}
where $G$ is the number of WiggleZ galaxies used as sphere centres, $R$ is the number of random catalogues, $N(<r)$ is the counts for WiggleZ galaxies, $N_{R}(<r)$ the counts for random galaxies (centred on the position of the $i^{\rm th}$ WiggleZ galaxy), and $\rho_j \equiv n_W/n_{\rm{rand},j}$ is the ratio of the total number of WiggleZ galaxies ($n_W$) to the number of random galaxies in the $j^{\rm th}$ random catalogue ($n_{\rm{rand},j}$). In our analysis, we have $G=n_W$, but this would not be the case if, for example, we excluded spheres near the survey edges.

The random catalogue correction has the effect of reducing the scaling by the number of dimensions (i.e. $D$=3), so that for a homogeneous distribution ${\mathcal{N}(<r)}$ scales as
\begin{equation}
\label{n_scaling}
\mathcal{N}(<r) \propto r^{3-3} = 1.
\end{equation}

\begin{figure}
\includegraphics[width=9cm]{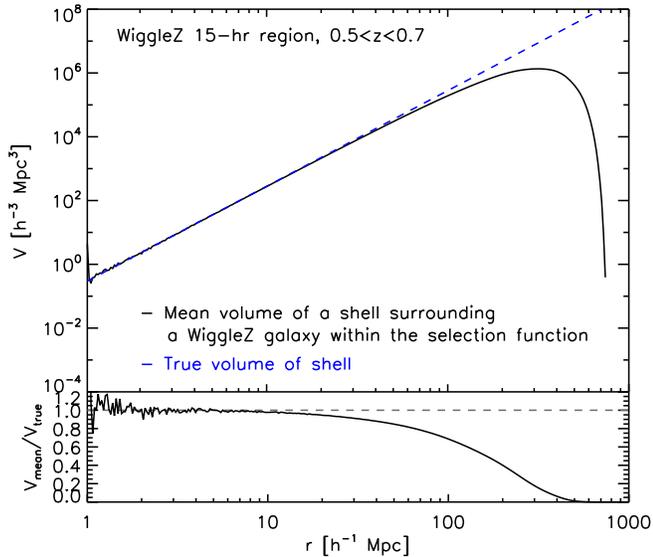}
\caption{ An illustration of the scales for which survey edge effects become important for the homogeneity measurement. Top panel:  The mean volume $\overline{V}(r)$ in a thin shell of mean radius $r$ surrounding a WiggleZ galaxy within the selection function (black curve). We show this for the 15-hr $0.5<z<0.7$ region. The true volume of the shells is shown as a blue dashed line. The black curve deviates from the blue at large scales, where an increasing proportion of the shells extends outside the survey. Bottom panel: The ratio of the mean volume to the true volume, as a function of $r$. The grey dashed line indicates a ratio of 1.} 
\label{sf_volume2}
\end{figure}

In each WiggleZ region we make ${\mathcal{N}(<r)}$ measurements in spheres with 12 to 15 logarithmically-spaced radial bins (depending on the size of the region). We determine the large-scale cutoff in each region by calculating the mean volume in the selection function, $\overline{V}(r)$, in a thin shell of mean radius $r$ enclosing a central galaxy. We illustrate this for the 15-hr ${0.5<z<0.7}$ region in Fig. \ref{sf_volume2}, and compare it with the `true' volume of the shells. Below $\sim2h^{-1}$Mpc there is noise due to low resolution, but above this the two curves are very close. On large scales however, an increasing proportion of shells surrounding galaxies go off the edge of the survey, and their volume within the survey decreases. On such scales, corrections for edge effects will become important. We take the large-scale cutoff of our homogeneity measurement at the radius of the maximum-volume shell. For the region in Fig. \ref{sf_volume2} this corresponds to shells that are $\sim20$ per cent complete. We show later that edge effects up to this scale do not impact our homogeneity measurement.

Our correction method using random catalogues maximises the use of the data, and accounts for incompleteness and the fact that WiggleZ is not volume-limited. However, it can potentially bias our result towards detecting homogeneity, since it assumes homogeneity on the largest scales of the survey. It is equivalent to weighting each measurement by the volume of the sphere included within the survey, multiplied by an arbitrary mean density. So while the counts-in-spheres measurement should have the advantage of not assuming a mean density \citep{hogg2005}, our correction method means that we do. Therefore ${\mathcal{N}(<r)}$ should tend to 1 on the largest scales of the survey, regardless of whether homogeneity has been reached. However, for a distribution with homogeneity size smaller than the survey, ${\mathcal{N}(<r)}$ should reach 1, and remain at 1, for a range of scales smaller than the survey scale. We check the robustness of our method against effects of the selection function and correction method in Section \ref{robustness} and show that our analysis is robust against this potential source of systematic error out to scales far greater than the homogeneity scale we measure.

There are other correction methods used in the literature. There is the so-called `exclusion' or `deflation' method \citep[e.g.][]{coleman1992,pan2000,pan2002,syloslabini2009} which only considers central points that are surrounded by complete spheres within the survey. This therefore excludes as central points any galaxies within a certain distance from the survey edges. However, this does not make the best use of the data, since it excludes data and so reduces the volume of the sample.

There is also the so-called `angular correction' model, which has been shown to be more optimal, by using all the available data without introducing a bias due to edge corrections \citep{pan2002}. This corrects measurements in a sphere by the solid angle subtended by regions in the sphere that are outside the survey boundary. However, as they point out, this method is difficult to apply to surveys with a complicated geometry, especially if they contain holes, as WiggleZ does. Finally, another correction that minimises bias at the survey edges, but wastes little data, is the `Ripley' estimator \citep{ripley1977,martinez1998}, which corrects the measurement in a sphere by the \textit{area} of the sphere contained within the survey. Both the angular correction and the Ripley estimator assume isotropy of the samples. Due to the geometry of the WiggleZ survey, we choose to use a random catalogue correction, but make robustness tests to quantify any bias it may introduce to the results.

To demonstrate the robustness of our measurement against the method of correcting for the selection function, we compare our method to an analysis using only complete spheres, with and without correcting for incompleteness, in Section \ref{corrections}, and show we obtain consistent results.

\subsection{Correlation dimension $D_2(r)$}
Fractal dimensions can be used to describe the clustering of a point distribution. There exists a general family of dimensions $D_q$, the Minkowski-Bouligand dimensions, which describe the scaling of  counts in spheres centred on points \citep[see e.g.][for a review]{borgani1995, martinez2002}. To completely characterise the clustering of our galaxy distribution we would need to consider all the moments $q$ of the distribution (corresponding to combinations of $n$-point correlation functions). However, to identify the scale of homogeneity we consider only $D_2$, the `correlation dimension,' which quantifies the scaling behaviour of the two-point correlation function $\xi(r)$.

If we take a galaxy and count the number of other galaxies, ${N(<r)}$, within a distance $r$, then this quantity scales as

\begin{equation}
N(<r) \propto r^{D_2},
\end{equation}
where $D_2$ is the fractal dimension of the distribution.

From this, the correlation dimension is defined as:
\begin{equation}
\label{D2}
D_2(r) \equiv {d\ln N(<r) \over d\ln r}.
\end{equation}

Since we must correct each WiggleZ $N{(<r)}$ measurement for completeness, obtaining the scaled quantity ${\mathcal{N}(<r) \propto r^{D_2-3}}$, we must calculate $D_2(r)$ via
\begin{equation}
D_2(r) = {d\ln \mathcal{N}(<r) \over d\ln r} + 3.
\end{equation}

For a homogeneous distribution, $D_2=3$. If $D_2 < 3$ then the distribution has a scale-invariant, fractal (and so inhomogeneous) clustering pattern. If $D_2>3$ the distribution is said to be `super-homogeneous' and corresponds to a lattice-like distribution \citep{gabrielli2002}. \color{black} A power-law in $1+\xi(r)\sim r^{-\gamma}$ has $D_2=3-\gamma$ for $\xi \gg 1$. 
      
Some previous works have found $D_2(r)$ simply by fitting a straight line to a log-log plot of ${\mathcal{N}(<r)}$. This method can give a false indication of a fractal \citep{martinez1998}; calculating $D_2(r)$ explicitly gives a more reliable measurement.
\\
\\

We note that our estimator for ${\mathcal{N}(<r)}$, Eq. \ref{nr_eq}, is essentially equivalent to $1 + \bar{\xi}_g(r)$, where \citep{hamilton1992} 
\begin{equation} 
\bar{\xi}(r) = {3 \over r^3} \int_0^r x^2 \xi(x) dx.
\end{equation}
(This can be seen more clearly by rearrangement of the theoretical expression for ${\mathcal{N}(<r)}$ given by Eq. \ref{Nr_theory}).  Many measurements of $\xi_g(r)$ have been made with different galaxy surveys \citep[e.g.][]{hawkins2003,zehavi2005,blake2011b,beutler2011}. 
On small scales the correlation function is well described by a power law,
\begin{equation}
\xi(r) = \left( {r_0 \over r} \right) ^\gamma,
\end{equation}
where $r_0 \approx 5 h^{-1}$Mpc is the so-called clustering length, and $\gamma \approx 1.8$, depending on the galaxy population. On scales $\gtrsim 20 h^{-1}$Mpc however, the correlation function is observed to turn over, consistent with large-scale homogeneity.

However, the correlation function cannot be used to test large-scale homogeneity, since the way it is determined from surveys depends on the mean galaxy density. Determinations of $\xi(r)$ commonly use the Landy-Szalay \citep{landy1993} or Hamilton estimators \citep{hamilton1993},
which compare the galaxy clustering to that of random catalogues of the same mean density as the survey. 

Our $\mathcal{N}(<r)$ estimator also compares the data to random catalogues, since we must correct for the selection function. Therefore $\mathcal{N}(<r)$, like $\xi(r)$, does assume a mean density on the scale of the survey. However, our estimator is slightly different, as we correct each object separately rather than an average pair count. 

The correlation dimension $D_2(r)$, on the other hand, measures the \textit{scaling} of $\mathcal{N}(<r)$, which is \textit{not} affected by the assumption of the mean density. (This only affects the amplitude of $\mathcal{N}(<r)$). Deviations from a volume-limited sample, which require random-catalogue corrections, only cause second-order changes to $D_2(r)$, whilst they would be leading-order in the raw correlation function. Therefore, $D_2(r)$ is much more robust to both the assumed mean density and details of the selection function, making it the most reliable measure of homogeneity.

\subsection{Lognormal realisations and covariance matrix}
\label{lognormals}

Lognormal realisations \citep{coles1991} are an important tool for determining uncertainties in galaxy surveys. A lognormal random field is a type of non-Gaussian random field, which can be used to model the statistical properties of the galaxy distribution, and simulate datasets with an input power spectrum. We have used 100 such realisations, generated using an input $\Lambda$CDM power spectrum with the fiducial parameters listed in Section \ref{intro}. These are sampled with the survey selection function, to create 100 mock catalogues for each of the WiggleZ regions. We use these to calculate the full covariance and errors of our measurement. Jack-knife resampling does not permit enough independent regions within the survey volume to give a reliable estimate of the uncertainties.

We obtain ${\mathcal{N}(<r)}$ and $D_2(r)$ for each of the lognormal realisations in the same way as for the WiggleZ data. The covariance matrix between radial bins $i$ and $j$ is given by

\begin{equation}
C_{ij} = {1 \over n-1}  \sum_{l=1}^n   [ x_l(r_i) - \overline{x(r_i)} ]  [ x_l(r_j) - \overline{x(r_j)} ],
\end{equation}
where $x(r)$ is ${\mathcal{N}(<r)}$ or $D_2(r)$, the sum is over lognormal realisations $l$, $n$ is the total number of lognormal realisations and $\overline{x(r)} = {1 \over n} \sum_{l=1}^n x_l(r)$. The diagonal values $j=k$ give the variance, $\sigma^2$.

The correlation coefficient between bins $i$ and $j$ is given by
\begin{equation}
\label{rij}
r_{ij} = {C_{ij} \over \sqrt{C_{ii}C_{jj}}}.
\end{equation}

The correlation matrices for ${\mathcal{N}(<r)}$ and $D_2(r)$ are shown in Figures \ref{nr_corrcoeff_z05_07_notitle} and \ref{D2r_corrcoeff_z05_07_notitle} respectively, for the combined regions (see next section) in the $0.5<z<0.7$ redshift slice.

It is noticeable that the ${\mathcal{N}(<r)}$ measurement is more correlated than $D_2(r)$. This is because ${\mathcal{N}(<r)}$ is effectively an integral of $D_2(r)$, so its covariance is effectively a cumulative sum of that of $D_2(r)$. It can also be explained by the fact that $D_2(r)$ is the logarithmic slope of ${\mathcal{N}(<r)}$, so it does not depend on the correlations between widely separated ${\mathcal{N}(<r)}$ bins but rather the variations between neighbouring bins.

We note that the uncertainties calculated using lognormal realisations assume $\Lambda$CDM, and represent the variance we would expect to measure in a $\Lambda$CDM universe. It would not be possible to calculate uncertainties for a fractal universe, since a fractal has no defined cosmic variance as it has no defined mean density. However, calculating uncertainties this way gives a valid consistency check of $\Lambda$CDM.

\subsection{Combining WiggleZ regions}

In each of the 4 redshift slices, we combine the measurements from each of the 7 WiggleZ regions using inverse-variance weighting. The combined measurements $x_{\rm comb}$ are given by

\begin{equation}
x_{\rm comb}(i) = { \sum_n w_n(i) x_n(i) \over \sum_n w_n(i)},
\end{equation}
where $n$ are the 7 WiggleZ regions, $x_n(i)$ is the measurement in the $i^{\rm th}$ radial bin in the $n^{\rm th}$ region, and $w_n(r) = 1 / C_n(i,i)$ is a weight function. $C_n(i,j)$ is the covariance matrix of the $n^{\rm th}$ region, and the combined covariance matrix is calculated by

\begin{equation}
C_{\rm comb}(i,j) = { \sum_n C_n(i,j) w_n(i) w_n(j) \over \sum_n w_n(i) \sum_n w_n(j) }.
\end{equation}

\begin{figure}
\includegraphics[width=8cm]{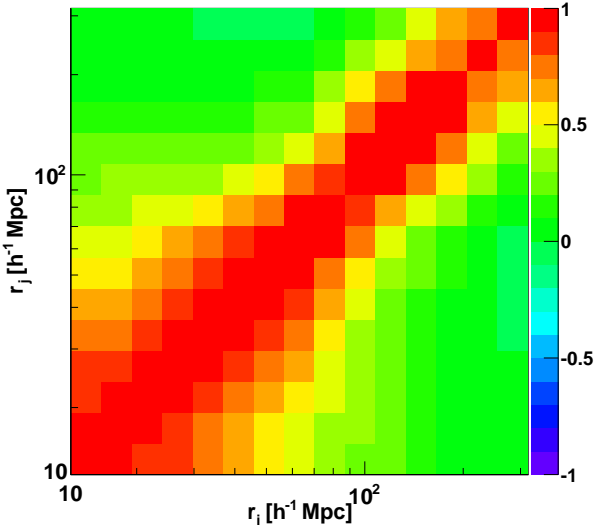}
\caption{The correlation matrix $C_{ij} / \sqrt{C_{ii}C_{jj}}$ for the $\mathcal{N}(r)$ measurement, obtained from lognormal realisations, for the ${0.5 < z < 0.7}$ redshift slice.}
\label{nr_corrcoeff_z05_07_notitle}
\end{figure}

\begin{figure}
\includegraphics[width=8cm]{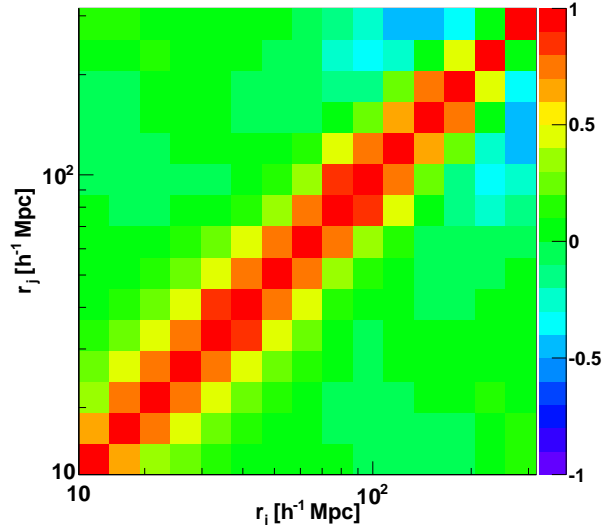}
\caption{The correlation matrix $C_{ij} / \sqrt{C_{ii}C_{jj}}$ for the $D_2(r)$ measurement, obtained from lognormal realisations, for the ${0.5 < z < 0.7}$ redshift slice.}
\label{D2r_corrcoeff_z05_07_notitle}
\end{figure}

We show $C_{\rm comb}(i,j)$ for the ${\mathcal{N}(<r)}$ and $D_2(r)$ measurements in the $0.5<z<0.7$ redshift slice, in terms of the correlation matrix $C_{ij} / \sqrt{C_{ii}C_{jj}}$, in Figs. \ref{nr_corrcoeff_z05_07_notitle} and \ref{D2r_corrcoeff_z05_07_notitle}.

\subsection{Likelihood ($\chi^2$) \& parameter fitting}

The $\chi^2$ of our model fit to the data is given by
\begin{equation}
\chi^2 \equiv \sum_{i=1}^{n_{\rm bins}} \sum_{j=1}^{n_{\rm bins}}  [x_{\rm th}(r_i) - x_{\rm obs}(r_i)] C^{-1}_{ij} [x_{\rm th}(r_j) - x_{\rm obs}(r_j)],
\end{equation}
where $x(r)$ is ${\mathcal{N}(<r)}$ or $D_2(r)$, $n_{\rm bins}$ is the number of radial bins, $x_{\rm th}$ is the theoretical model and $x_{\rm obs}$ is the measured value. $C_{ij}^{-1}$ is the inverse of the covariance matrix. 

\subsection{A new method of defining the `homogeneity scale' $R_H$}
\label{rh}
As previously mentioned, the definition of a `homogeneity scale' is somewhat arbitrary since we expect the Universe to have a gradual approach to homogeneity. Recently \cite{bagla2008} and \cite{yadav2010} proposed that the homogeneity scale be defined where the measured fractal dimension $D_q$ becomes consistent with the ambient dimension within the 1$\sigma$ statistical uncertainty, $\sigma_{\Delta D_q}$. They derived an approximation for $D_q(r)$ and $\sigma_{\Delta D_q}$, in the limit of weak clustering, for a given correlation function, and showed that both scale the same way, and so the homogeneity scale stays constant, with bias and epoch. This definition is therefore beneficial as it is not arbitrary, and is robust to the tracer galaxy population. However, in deriving $\sigma_{\Delta D_q}$ they considered only shot noise, and cosmic variance from variance in the correlation function, while ignoring contributions from the survey geometry and selection function. Any real survey will have these contributions to the statistical uncertainty, which cannot be separated from the variance due to the correlation function alone. The value derived from this definition in a real survey is therefore difficult to interpret in a meaningful way, i.e. one that allows comparison with theory, or with different surveys of differing volume and selection function.

\begin{figure}
\includegraphics[width=9cm]{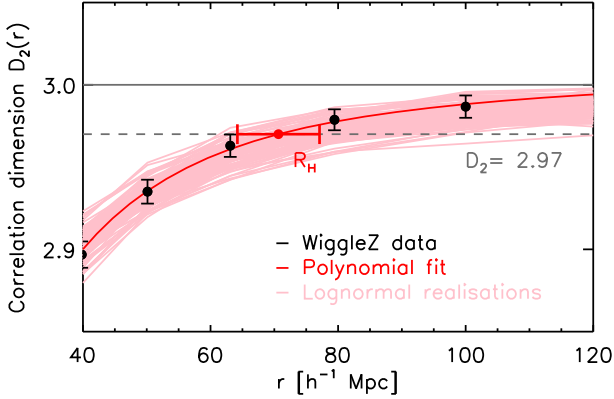}
\caption{Illustration of our method of defining the homogeneity scale, $R_H$, shown here for the $D_2(r)$ measurement. We first fit a model-independent polynomial (red curve) to the data (black data points). We then find where this intercepts a chosen value close to homogenenity, e.g. 1 per cent from homogeneity, $D_2=2.97$ (dotted grey line). This gives us $R_H$. We find the uncertainty in $R_H$ from the root mean square variance of 100 lognormal realisations (pink curves).}
\label{RH_cartoon}
\end{figure}

We therefore introduce a different method for determining a `homogeneity scale' $R_H$, which is easier to compare with theory and between surveys. Our method is to fit a smooth, model-independent polynomial to the data, and find the scale at which this intercepts a chosen value, or `threshold,' close to homogeneity. This scale is then defined as the homogeneity scale $R_H$. For example, $R_H$ could be the value of $r$ at which the polynomial intercepts a line 1 per cent from $\mathcal{N}(<r)=1$, or $D_2(r)=3$ (see Fig. \ref{RH_cartoon} for an illustration). The uncertainty is found using the 100 lognormal realisations. The homogeneity scale measured this way does not depend directly on the survey errors (although the \textit{uncertainties} on $R_H$ do), and is less susceptible to noise in the data, making this a preferable method that allows comparisons between different surveys. It also allows easy comparison between the data and a given model, e.g. $\Lambda$CDM, and we can check that the data converges to $\mathcal{N}=1$ or $D_2=3$ as expected for a homogeneous distribution, by choosing a range of thresholds approaching homogeneity (see Fig. \ref{Rh_vs_b2sigma82_3}).

We can also take this further and construct a likelihood distribution for the homogeneity scale, as described in Section \ref{likelihood}.

Although our choice of $D_2$ threshold is arbitrary, by choosing the same threshold for different surveys we obtain an $R_H$ value that can be meaningfully compared, and that can be easily compared to a theoretical model. Our choice of threshold may be limited by the amount of noise in the data, however. For instance, we can measure an intercept 1 per cent away from homogeneity for the WiggleZ data, but cannot measure 0.1 per cent in two of the redshift slices, due to noise (the data do not come this close to homogeneity, although they are consistent with it within the uncertainties) -- see Fig. \ref{Rh_vs_b2sigma82_3}. Nonetheless, we can easily choose a threshold that is possible given our data, and use this to compare with a model. The Baryon Acoustic Oscillation (BAO) feature can also potentially affect the appropriate choice of intercept value as it causes a small distortion in $D_2(r)$ -- we discuss this in Section \ref{discussion}.

 For the main results in this paper we have chosen a threshold of 1 per cent away from homogeneity, since this is about the closest threshold to homogeneity we can measure, considering the noisiness of the data.

\section{$\Lambda$CDM model prediction of ${\mathcal{N}(<r)}$ and $D_2(r)$}
\label{analytic}

In this section we derive theoretical $\Lambda$CDM predictions for the counts-in-spheres and correlation dimension. This allows us to compare our measurements of the transition to homogeneity to the predictions of a $\Lambda$CDM model that fits WMAP data.

\subsection{${\mathcal{N}(<r)}$ and $D_2(r)$}

For a particular galaxy population, we can calculate the mean counts-in-spheres ${N(<r)}$ from the 2-point matter correlation function predicted by $\Lambda$CDM. The 2-point correlation function $\xi(r)$ is defined as the excess probability above random of finding two objects in volumes $dV_1$ and $dV_2$, separated by distance $r$ \citep{peebles1980}:
\begin{equation}
P(r) = \bar{\rho}^2 [1+\xi(r)] dV_1 dV_2,
\end{equation}
where $\bar{\rho}$ is the mean number density. 

The mean number of galaxies surrounding a random galaxy up to distance $r$ is found by integrating the correlation function
\begin{equation}
\label{nr_analytic}
N(<r) = \bar{\rho} \int_0^r [1+b^2 \xi(r')] 4 \pi r'^2 dr',
\end{equation}
where $b$ is the galaxy bias, relating the clustering of a particular galaxy population to the underlying dark matter distribution. Note that $\Lambda$CDM assumes large-scale homogeneity, and indeed we must assume large-scale homogeneity in order for a mean density $\bar{\rho}$ to be defined.

We obtain our model correlation function by transforming a  $\Lambda$CDM matter power spectrum $P_{\delta\delta}(k)$ generated using CAMB \citep{lewis2000}. Since we make our measurements in redshift space, we first convert $P_{\delta\delta}(k)$ to the redshift-space galaxy power spectrum $P_g^s(k)$ (described in the next section), then convert this to the redshift-space galaxy correlation function $\xi_g(s)$, where $s$ denotes distance in redshift-space. Since we use the angle-averaged power spectrum (assuming the power spectrum is isotropic), we do not need to integrate the angular part of the $k$-space integral, and so use a spherical Hankel transform rather than a Fourier transform to obtain $\xi_g(s)$:  

\begin{equation}
\xi_g(s) = {1 \over 2 \pi^2 } \int  P_g^s(k) {\sin{ks} \over ks} k^2 dk.
\end{equation}

To compare with our WiggleZ measurement (Eq. \ref{nr_eq}), where we correct for incompleteness, we divide our counts-in-spheres prediction by the number that would be expected for a random distribution, i.e. $\bar{\rho} {4 \over3} \pi r^3$:
\begin{equation}
\label{Nr_theory}
\mathcal{N}(<r) = {3 \over 4 \pi r^3} \int _0 ^r [1 + \xi_g(s)]4\pi s^2 ds.
\end{equation}

We calculate the model $D_2(r)$ by simply applying Eq. \ref{D2} to our model $\mathcal{N}(<r)$.

\subsection{Redshift-space distortions and nonlinear velocity damping}
\label{RSDs}

Here we describe how we implement redshift-space distortions in our analytical model. In practice, we measure the positions of galaxies in redshift-space, which are affected by redshift-space distortions. These are due to the peculiar velocities of galaxies along the line of sight, which add to the measured redshifts and perturb the inferred galaxy positions. This anisotropic effect creates anisotropy in the observed redshift-space galaxy power spectrum $P^s_g(k,\mu)$, and can be modelled by multiplying (convolving in  configuration space) the real-space matter power spectrum by an angle-dependent function $F(k,\mu)$:
\begin{equation}
P^s_g(k,\mu) = b^2 F(k,\mu)P_{\delta\delta}(k).
\end{equation}

There are two forms of redshift-space distortion of relevance to our measurement, which we find are necessary for a good fit to the data:

\begin{enumerate}
\item On large, linear scales ($\gtrsim 20 h^{-1}$ Mpc) the bulk infall of galaxies towards overdensities creates an enhancement in the observed power spectrum along the line of sight. This can be modelled by the linear Kaiser formula \citep{kaiser1987}, 
	\begin{eqnarray}
	P^s_g(k,\mu) &=& b^2 \left(1+{f \mu^2 \over b} \right) ^2P_{\delta\delta,L}(k) \nonumber \\
	&=& b^2 (1+\beta\mu^2)^2P_{\delta\delta,L}(k),
	\end{eqnarray}
where $P_{\delta\delta,L}(k)$ is the linear matter power spectrum, $f$ is the growth rate of structure and $\beta=f/b$ is the redshift-space distortion parameter.

Note that this formula assumes a perturbed FRW universe with small real-space density perturbations $|\delta(r)| \ll 1$. 

\item On quasilinear scales (${10 \lesssim s \lesssim 20 h^{-1}}$ Mpc ) the peculiar velocities resulting from the scale-dependent growth of structure distort the shape of the power spectrum, via a scale-dependent damping effect. A common way of modelling this is the `streaming model' \citep{peebles1980, fisher1995,hatton1998}, which combines the linear theory Kaiser formula with a velocity streaming term.  We choose to use a Lorentzian term, ${F=[1+(k\sigma_p\mu)^2]^{-1}}$, for an exponential velocity probability distribution function, since \cite{blake2011a} found this to be a good fit to the WiggleZ power spectrum for $k<0.1h^{-1}$Mpc. This gives us the so-called `dispersion model' \citep{peacock1994} for the full redshift-space power spectrum: 
\begin{equation}
\label{pgskmu}
P^s_g(k,\mu) = b^2{(1+\beta\mu^2)^2 \over 1+(k\sigma_p\mu)^2} P_{\delta\delta}(k).
\end{equation}
Here, $\sigma_p$ is the pairwise velocity dispersion along the line of sight. Both $\sigma_p$ and $\beta$ are parameters that must be fitted to the data. We use the values obtained by \cite{blake2011a} in each redshift slice from fits to the WiggleZ two-dimensional power spectrum -- these are listed in Table \ref{betasigmav}.
	
We note that the streaming model is motivated by virialised motions of particles within haloes, on much smaller scales - the so-called `Finger of God' effect at $\lesssim 2 h^{-1}$ Mpc. However, it is heuristic in nature and can also describe physical scales of tens of $h^{-1}$Mpc. \cite{blake2011a} apply it this way by fitting for $\sigma_p$ on these scales, rather than on Finger-of-God scales. We find that including it gives a significant improvement of our model fit to data.

\end{enumerate}

 \begin{table}
 \begin{center}
 \caption{Values of the redshift space distortion parameter $\beta$, and the pairwise velocity dispersion $\sigma_p$, used in our modelling of nonlinear redshift-space distortion effects. These values were obtained by {\protect \cite{blake2011a}} from fits to the WiggleZ two-dimensional power spectrum.}
 \label{betasigmav}
 \begin{tabular}{@{}ccccccc}
  \hline
Redshift & $\beta$ & $\sigma_p$ [$h$ km/s]  \\
  \hline
  $0.1<z<0.3$ & 0.69 &  346 \\
$0.3<z<0.5$ & 0.73 & 275 \\
 $0.5<z<0.7$ & 0.60& 275 \\
$0.7<z<0.9$ & 0.51  &  86 \\
  \hline
 \end{tabular}
  \end{center}
 \end{table}

To obtain the angle-averaged redshift-space galaxy power spectrum $P_g^s(k)$ we need to convert the full $F(k,\mu)$ to $F(k)$, which we do by integrating over $\mu$:
\begin{equation}
F(k) = \int_{\mu=0}^1 {(1+\beta\mu^2)^2 \over (1+(k\sigma_v\mu)^2)} d\mu .
\end{equation}

The angle-averaged redshift-space galaxy power spectrum is then
\begin{equation}
P_g^s(k) = b^2 F(k)P_{\delta\delta}(k).
\end{equation}

Eq. \ref{pgskmu} normally assumes a linear matter power spectrum; however, we choose to use a nonlinear $P_{\delta\delta}(k)$, calculated by CAMB using the HALOFIT code \citep{smith2003}, since \cite{blake2011a} found this gave a better fit to the data.

\subsection{Correcting the WiggleZ data for galaxy bias}
\label{correcting}

The amplitude of the galaxy correlation function is affected by galaxy bias and redshift-space distortions, and the shape is affected by nonlinear damping. Therefore, these also affect the amplitude and shape of ${\mathcal{N}(<r)}$, as well as the measured scale of homogeneity.  It is possible to correct the data for bias, and so determine the homogeneity scale for the underlying matter distribution, by assuming a particular model, i.e. our $\Lambda$CDM + WMAP model, fitting for bias by minimising the $\chi^2$ value, and correcting the data for this. 

In our analysis, we only consider linear galaxy bias. This relates the galaxy correlation function $\xi_g(r)$ to the matter correlation function $\xi_m(r)$ through $\xi_g(r) = b^2 \xi_m(r)$. Since we are only interested in large scales, we do not consider scale-dependent bias that may occur on small scales.

We fix the redshift space distortion parameter $\beta$ and velocity dispersion $\sigma_p$ to the values listed in Table \ref{betasigmav}.  We then obtain our corrected measurements $\mathcal{N}_{\rm biasfree}(<r)$, 
\begin{equation}
\mathcal{N}_{\rm biasfree}(<r) = {\mathcal{N}(<r)-1 \over b^2} + 1,
\end{equation}
and we calculate the bias-corrected correlation dimension $D_{\rm 2, biasfree}(r)$ from this, using Eq. \ref{D2}.

Since we assume a $\Lambda$CDM model to fit for bias, we cannot make a model-independent measurement of transition to homogeneity of the underlying matter distribution. However, it is still interesting to look at the variation of the homogeneity scale of the matter distribution with redshift, assuming $\Lambda$CDM + WMAP.
\\
\\
We note that our measurement of the homogeneity scale of the WiggleZ galaxies is independent of the $\Lambda$CDM modelling shown in this section. This modelling is only done so that we can show that the measurement is consistent with that expected from $\Lambda$CDM.

\section{Results}
\label{results}

\subsection{${\mathcal{N}(<r)}$ and $D_2(r)$}

\begin{figure*}
\includegraphics[width=18cm]{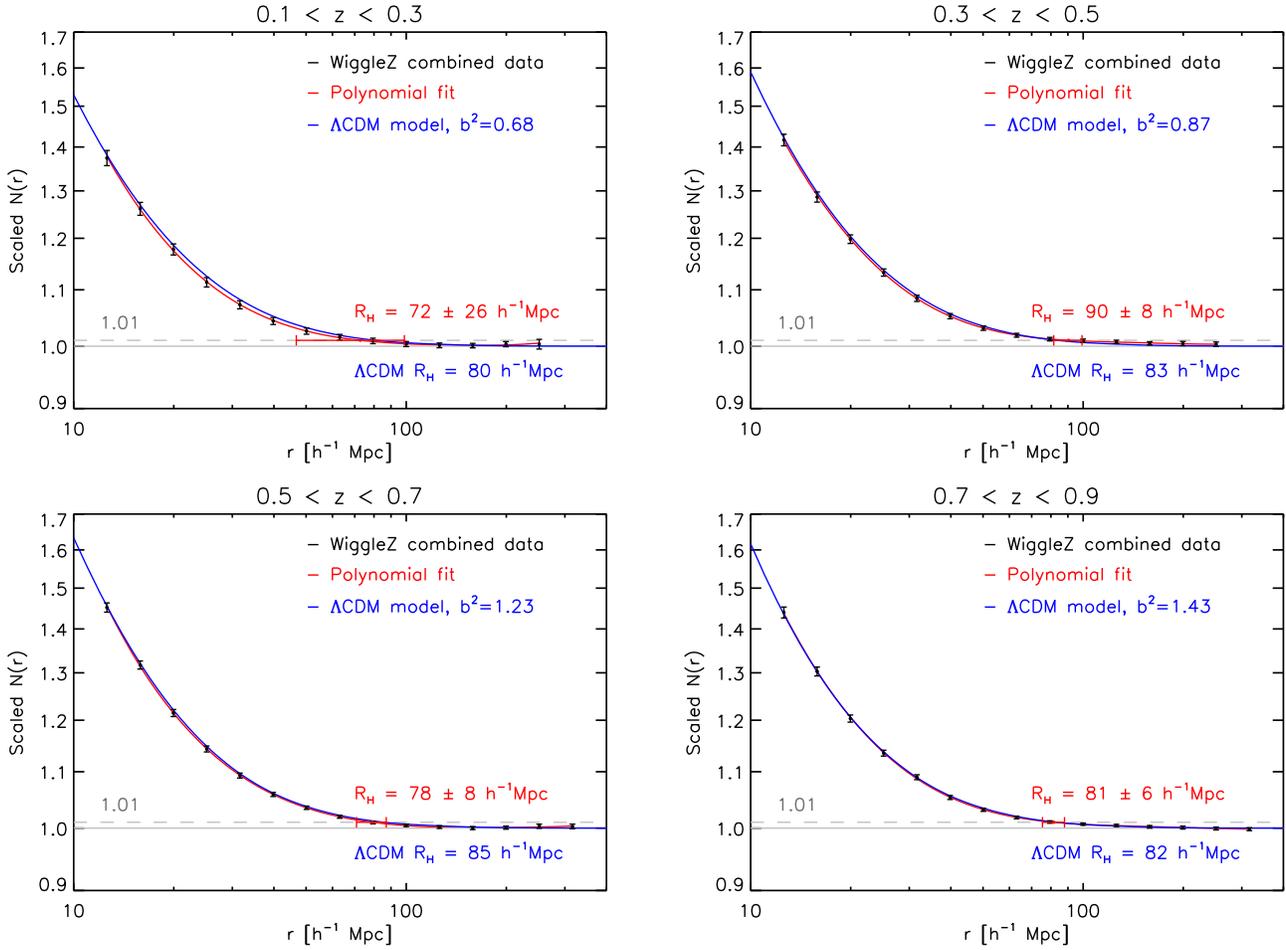}
\caption{Scaled counts-in-spheres ${\mathcal{N}(<r)}$ for the combined WiggleZ data in each of the four redshift slices (black error bars). A $\Lambda$CDM model with best-fitting bias $b^2$ is shown in blue. A 5$^{\rm th}$-degree polynomial fit to the data is shown in red. The red errorbar and label show the homogeneity scale $R_H$ for the galaxy distribution, measured by the intercept of the polynomial fit with 1.01 (1 per cent away from homogeneity), with the error given by lognormal realisations. This scale is consistent with the $\Lambda$CDM intercept with 1.01, labelled in blue.}
\label{nr_fit_101title}
\end{figure*}

The ${\mathcal{N}(<r)}$ measurements in each of the four redshift slices are shown in Fig. \ref{nr_fit_101title}. The data are compared with a $\Lambda$CDM+WMAP model (described in Section \ref{analytic}). For each successive redshift slice, the reduced $\chi^2$ values are 0.57, 0.91, 0.69 and 1.1. The first two redshift slices have 14 data bins from 12.5 to 251 $h^{-1}$Mpc, while the last two have 15 data bins from 12.5 to 316 $h^{-1}$Mpc. The data is consistent with a monotonically decreasing function, so we can fit a polynomial and find where this intercepts a chosen threshold, as per our definition of homogeneity.

The intercepts of the polynomial fit with $\mathcal{N}=1.01$ (1 per cent away from homogeneity), which we define as the homogeneity scale $R_H$,  are shown as red error bars. The errors were determined from the 100 lognormal realisations, and correspond to the square-roots of the diagonal elements of their covariance matrix.  The $R_H$ values and their errors are shown in Fig. \ref{nr_fit_101title} and listed in Table \ref{resultstable}, along with the values for the $\Lambda$CDM model, which are in good agreement.

\begin{table*}
\caption{Measured values of  the homogeneity scale $R_H$ (where data intercepts 1 per cent of the homogeneity value). The $R_H$ values shown are all for the galaxy distribution, except the values for the bias-corrected data, which are for the underlying matter distribution, i.e. $b^2=1$. The bias-corrected values directly assume a $\Lambda$CDM+WMAP model.  }
 \label{resultstable}
 \begin{tabular}{@{}cccccccc}
  \hline
  & WiggleZ data & $\Lambda$CDM & Measured & Bias-corrected & $\Lambda$CDM,  & Likelihood analysis & Bias-corrected  \\
   & [Mpc/h] & [Mpc/h] &bias, $b^2$ & data [Mpc/h] & $b^2=1$ [Mpc/h] & [Mpc/h] & likelihood analysis [Mpc/h] \\
  \hline
  ${\mathcal{N}(<r)}$ \\
$0.1<z<0.3$ & $72 \pm 26$ & 80 & 0.68 & $83 \pm 35$  &93 & - & -    \\
$0.3<z<0.5$ & $90 \pm 8$ & 83 & 0.87 & $99 \pm 11$  &87 & - & -   \\
$0.5<z<0.7$ & $78 \pm 8$& 85 & 1.23 & $73 \pm 6$  &79 & - & -   \\
$0.7<z<0.9$ & $81 \pm 6$& 82 & 1.43 &  $70 \pm 4$  &72 & - & -   \\
\hline
 $D_2(r)$ \\
$0.1<z<0.3$ & $70 \pm 7$ & 76 & 0.70 & $84\pm11$  &88 & $71\pm8$ &  $89\pm14$   \\
$0.3<z<0.5$ & $70 \pm 5$ & 78 & 0.87 & $74\pm6$  &82 & $70\pm5$ & $75\pm5$  \\
$0.5<z<0.7$ & $81 \pm 4$ & 81 & 1.25 & $74\pm3$  &73 & $81\pm 5$ & $76\pm4$ \\
$0.7<z<0.9$ & $74 \pm 4$ & 78 & 1.46 & $64\pm2$  &66 & $75\pm4$ & $65\pm4$ \\
 \hline

 \end{tabular}
 \end{table*}

\begin{figure*}
\includegraphics[width=18cm]{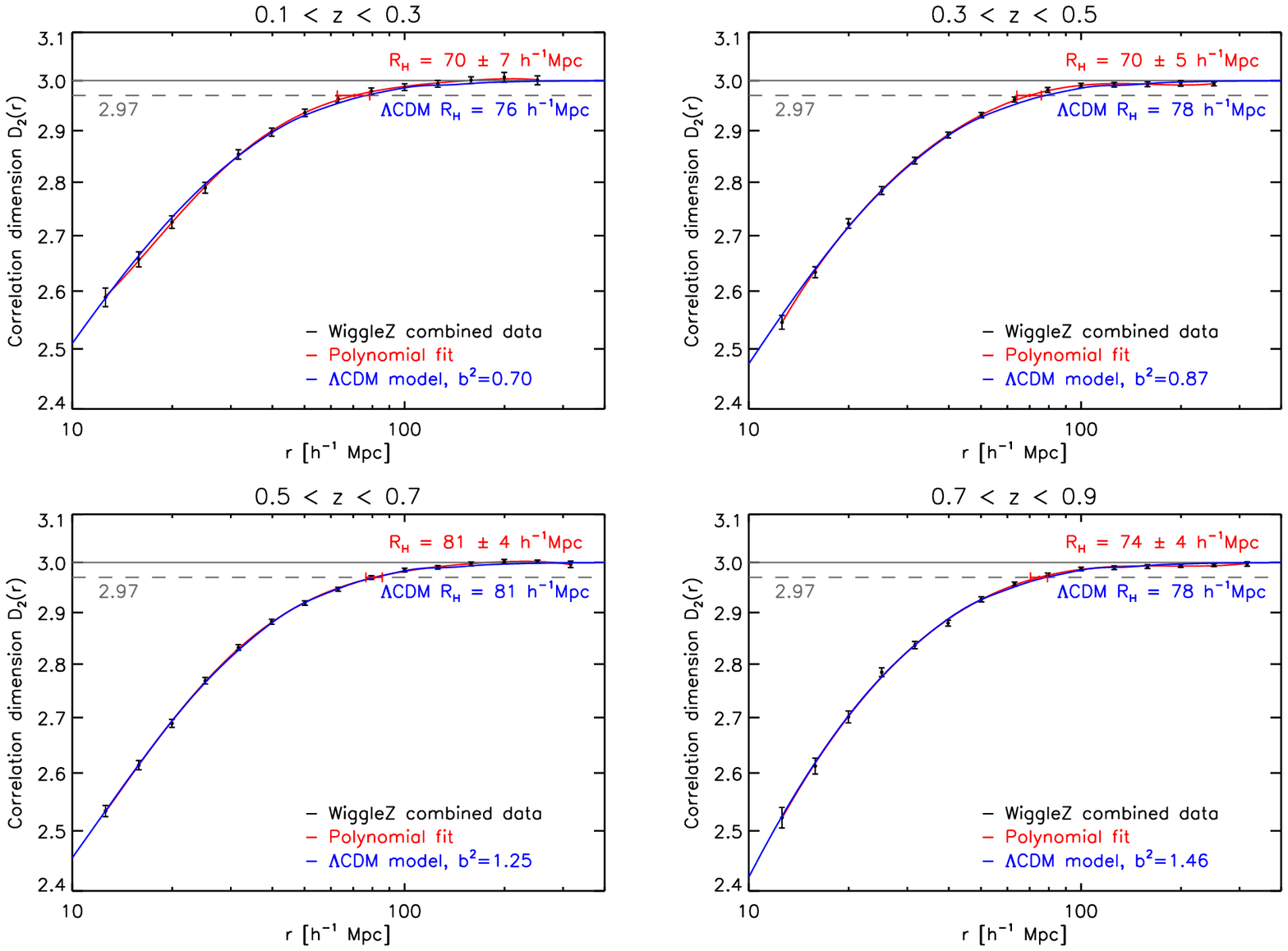}
\caption{Same as for Fig. \ref{nr_fit_101title} but for the correlation dimension $D_2(r)$. The $D_2(r)$ measurements for the combined WiggleZ data in each of the four redshift slices are shown as black error bars. A $\Lambda$CDM model with best-fitting bias $b^2$ is shown in blue. A 5th-degree polynomial fit to the data is shown in red. The red errorbar and label show the homogeneity scale $R_H$ measured by the intercept of the polynomial fit with 2.97 (1 per cent away from homogeneity), with the error given by lognormal realisations. This scale is consistent with the $\Lambda$CDM intercept with 2.97, labelled in blue.}
\label{D2_fit_299title}
\end{figure*}

The $D_2(r)$ measurements in each of the redshift slices are shown in Fig. \ref{D2_fit_299title}, along with a $\Lambda$CDM+WMAP model with best-fitting bias. In each redshift slice the data range and degrees of freedom are the same as for the ${\mathcal{N}(<r)}$ measurement. The reduced $\chi^2$ values in each redshift slice are 0.83, 0.90, 0.74 and 0.98. In each case a polynomial is fitted to the data. The homogeneity scales measured where these intercept 1 per cent away from homogeneity, $D_2=2.97$, are listed in Table \ref{resultstable} and are also in excellent agreement with the $\Lambda$CDM values.

\subsection{Effect of bias and $\sigma_8(z)$ on $R_H$}

The homogeneity scale of the model galaxy distribution, measured at 1 per cent from homogeneity, will depend on the amplitude and shape of the correlation function, and so on galaxy bias $b$ and $\sigma_8(z)$. We would also expect it to depend on redshift, since $\sigma_8(z)$ increases over time. These two parameters are in fact completely degenerate in the ${\mathcal{N}(<r)}$ and $D_2(r)$ measurements. 

\begin{figure*}
\includegraphics[width=18cm]{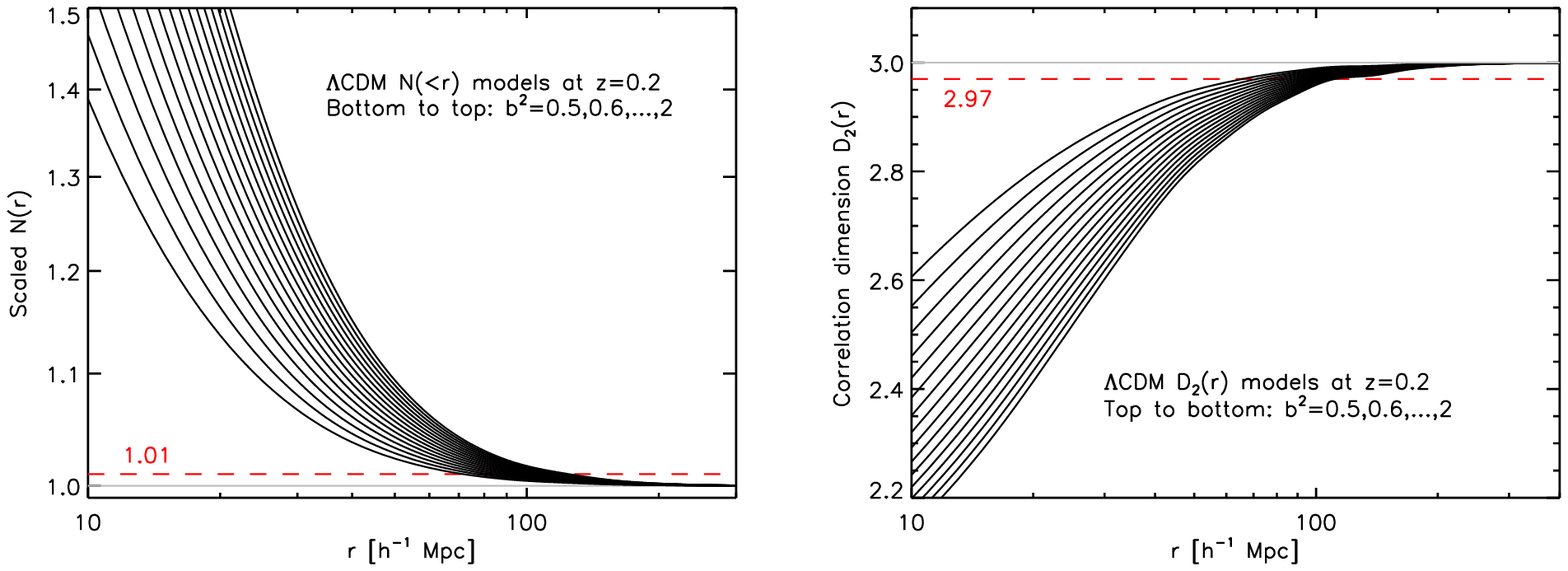}
\caption{The effect of bias on a $\Lambda$CDM ${\mathcal{N}(<r)}$ model (left) and $D_2(r)$ model (right) at $z=0.2$. Increasing bias increases the value of ${\mathcal{N}(<r)}$ on small scales, and decreases the value of $D_2(r)$ on small scales,  and produces a larger homogeneity scale, as seen by the intercepts of the curves with 1 per cent of homogeneity ($\mathcal{N}=1.01$ and $D_2=2.97$, red dotted lines). }
\label{analytic_nr_D2}
\end{figure*}

Fig. \ref{analytic_nr_D2} shows how our $\Lambda$CDM ${\mathcal{N}(<r)}$ and $D_2(r)$ models vary with bias, at $z=0.2$, for fixed $\sigma_8(z=0)=0.8$. Larger bias means a larger amplitude of clustering, so that both curves are steeper on small scales. This means that the models reach homogeneity at larger radii for higher bias. This can be understood qualitatively, since highly biased galaxies are more clustered together than less biased galaxies, so we must go to larger scales before we reach a homogeneous distribution.

\begin{figure}
\includegraphics[width=9cm]{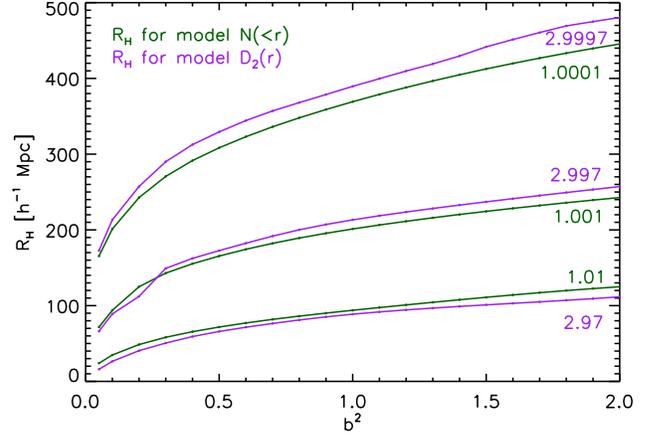}
\caption{$R_H$ for $\Lambda$CDM $\mathcal{N}(r)$ (green) and $D_2(r)$ (purple) models at $z=0.2$ with differing bias $b^2$. Each curve corresponds to $R_H$ evaluated at a different threshold -- 1, 0.1 or 0.01 per cent away from homogeneity (from bottom to top, labelled).}
\label{Rh_vs_b}
\end{figure}

Fig. \ref{Rh_vs_b} shows how the homogeneity scale $R_H$ of the galaxy distribution varies with bias, for different intercept values approaching homogeneity, for the ${\mathcal{N}(<r)}$ and $D_2(r)$ models at $z=0.2$ with fixed $\sigma_8(z=0)=0.8$. For a particular intercept value, larger bias again gives a larger homogeneity scale. Since ${\mathcal{N}(<r)}$ and $D_2(r)$ approach homogeneity asymptotically, the intercept jumps to higher values when we consider an intercept value closer to homogeneity. This image also shows the mapping between ${\mathcal{N}(<r)}$ and $D_2(r)$ homogeneity values for intercepts at 1, 0.1 and 0.001 percent away from homogeneity. They are not identical since they are slightly different methods, but give similar results. This plot illustrates that there are many potential ways to define homogeneity, and so it is important to make consistent  measurements between surveys in order for them to be comparable with each other and with theory.

Our bias-corrected ${\mathcal{N}(<r)}$ and $D_2(r)$ measurements are listed in Table \ref{resultstable}. The errors on the bias-corrected data were determined by applying a bias correction to each of the 100 lognormal realisations individually, and recalculating the covariance matrix. Since the bias correction aims to set the bias of all the realisations to $b=1$, it lowers the overall variance, and so the error bars are slightly smaller than for the pure data. These measurements give our measured homogeneity scale for the matter distribution, assuming $\Lambda$CDM. We find that this scale increases with redshift, as expected in $\Lambda$CDM. However, since we are assuming $\Lambda$CDM, which has $\sigma_8(z)$ increasing with time, this is not a model-independent result.

\begin{figure*}
\includegraphics[width=18cm]{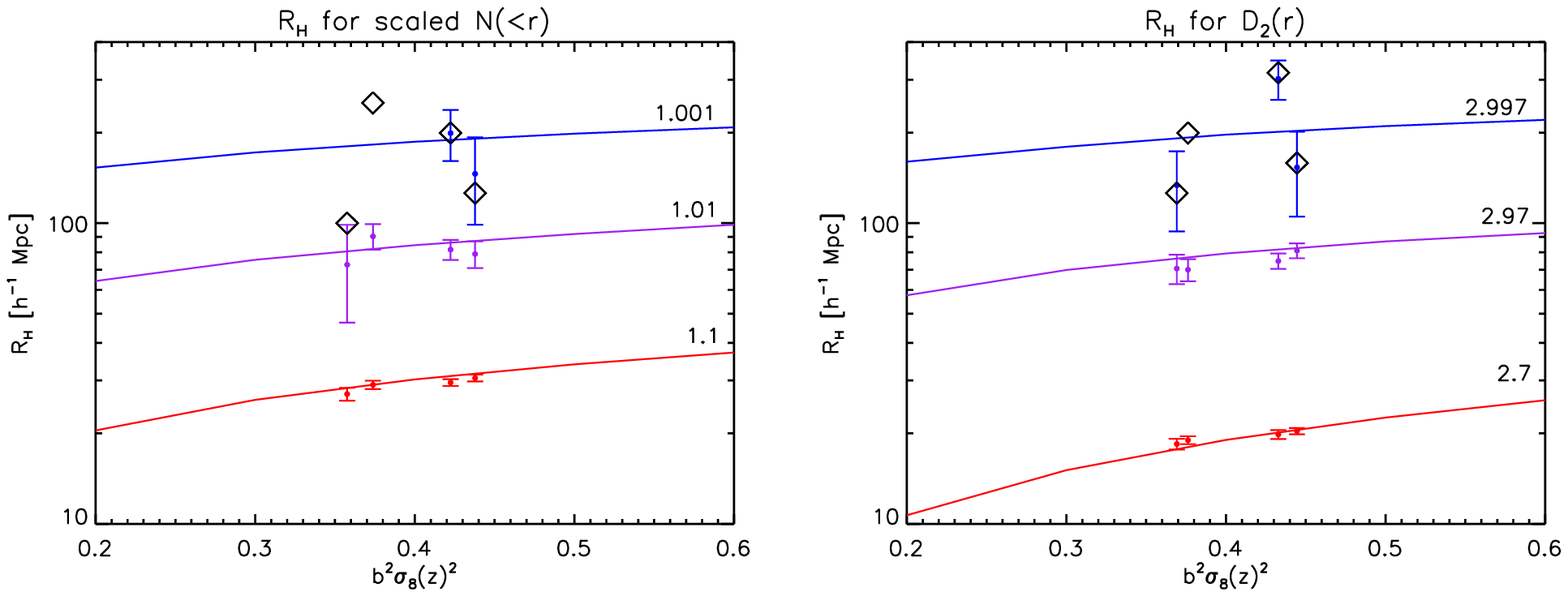}
\caption{Homogeneity scale $R_H$ as a function of $b^2\sigma_8(z)^2$, as predicted by the $\Lambda$CDM model, for different thresholds approaching homogeneity (10, 1 and 0.1 per cent from homogeneity, coloured curves from bottom to top), for ${\mathcal{N}(<r)}$ (left) and $D_2(r)$ (right). The corresponding WiggleZ results are shown as error bars of corresponding colour (the errors are found using lognormal realisations). The $b^2\sigma_8(z)^2$ values of the data increase with redshift slice (so the data points from left to right are from low to high redshift). Not all redshift slices have measurements at 0.1 per cent (blue) since the data do not reach this value in those slices. The definition of homogeneity used by previous authors is where the data comes within $1 \sigma$ of homogeneity -- we show these scales as black diamonds. It is clear that this definition has much greater stochasticity than our definition, which fits a smooth curve to many data points. Indeed, this approach gives quite uninformative results in this plane and cannot be compared to the model prediction.}
\label{Rh_vs_b2sigma82_3}
\end{figure*}

As already mentioned, the effect of bias on the homogeneity scale of galaxies is degenerate with the amplitude of the correlation function, $\sigma_8(z)$, since the correlation function at redshift $z$ depends on a combination of these, ${\xi(r,z) \propto b^2\sigma_8(z)^2}$. So far we have assumed a fixed value of $\sigma_8(z=0)$. But we can also make predictions independent of $\sigma_8$, by finding how $R_H$ changes as a function of the combination $b^2\sigma_8(z)^2$. This is shown in Fig. \ref{Rh_vs_b2sigma82_3}. We also show the WiggleZ results, which we have plotted for the best-fit $b^2\sigma_8(z)^2$ value in each redshift slice. This increases with redshift, so the data points from left to right go from low to high redshift. The WiggleZ results are in very good agreement with the $\Lambda$CDM+WMAP predictions. We show, for comparison, the values obtained from defining the homogeneity scale as where the data comes within 1$\sigma$ of homogeneity. These values have much greater stochasticity than those from our method of fitting a smooth curve to many data points, and do not give informative results in this plane.

We see that the model $R_H$-$b^2\sigma_8(z)^2$ curves are monotonically increasing. Since we expect $\sigma_8(z)$ in $\Lambda$CDM to grow over time due to growth of structure, we would therefore also expect the homogeneity scale to increase over time, for galaxies with fixed bias. 

For the WiggleZ data, however, the measured homogeneity scale does not appear to decrease with redshift. This is explained by the fact that the WiggleZ galaxies have increasing bias with redshift, assuming a $\Lambda$CDM growth rate. As explained previously, this is understood to be due to the effects of Malmquist bias and downsizing on the selection of the WiggleZ galaxy population, and the colour and magnitude cuts. This counteracts the effect of decreasing $\sigma_8$ with redshift. As can be seen in Fig. \ref{Rh_vs_b2sigma82_3}, within the $b^2\sigma_8(z)^2$ range of the data we would not expect a significant change in $R_H$, measured at 1 per cent from homogeneity, with redshift.

\subsection{Likelihood analysis for homogeneity scale}
\label{likelihood}

Rather than trying to measure the scale of homogeneity directly, it can be informative to consider the likelihood that the Universe has reached homogeneity by a certain scale. We can construct a probability distribution function (PDF) for the homogeneity scale, by combining the likelihood distribution of the data with that of the definition of the homogeneity scale. This gives the probability that homogeneity is reached at a certain scale $r$.

We could apply a likelihood analysis to either our ${\mathcal{N}(<r)}$ or $D_2(r)$ measurements. However, it would arguably be invalid for highly correlated data, such as the ${\mathcal{N}(<r)}$ measurement, since the different contributions to the probability distribution would be correlated, and we do not correct for this. The $D_2(r)$ measurement is much less correlated (see Figs. \ref{nr_corrcoeff_z05_07_notitle} and \ref{D2r_corrcoeff_z05_07_notitle}). Also, as we have explained, $D_2(r)$ is the most robust measurement of homogeneity, so we have chosen this for our likelihood analysis. 

The likelihood distribution on the data is simply given by the mean and variance of the lognormal realisations in each bin. These give the expected variance of a $\Lambda$CDM distribution sampled with the WiggleZ selection function, so take into account both cosmic variance and shot noise. Here we assume they are Gaussian-distributed by virtue of the central limit theorem. However we also consider the true distributions provided by the lognormal realisations -- this gives similar results but with larger errors, see Appendix \ref{pdfappendix}. 

\begin{figure}
\includegraphics[width=9cm]{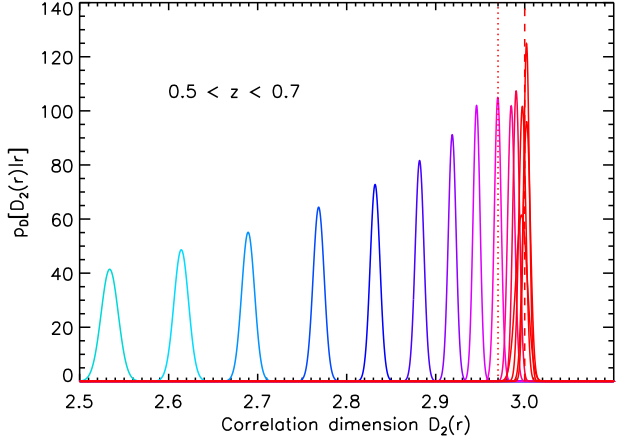}
\caption{The probability distributions $p_D[D_2(r)]$ for each of the $r$ bins in the $0.5<z<0.7$ redshift slice. Blue-to-red gradient indicates small to large radius, from 12.5 to 316 $h^{-1}$Mpc. The area of each distribution above $D_2(r)=2.97$ (dotted line) gives the probability that homogeneity has been reached within 1 per cent by that value of $r$.  }
\label{nr_pdfs_int1_D2_z57}
\end{figure}

At each $r$, the data therefore provide a probability distribution $p_D[D_2(r)]$, which we model as Gaussians, as:
\begin{equation}
p_D[D_2(r)] = {1 \over  \sigma\sqrt{2 \pi}} e^{-{(D_2(r)-\mu)^2 / 2\sigma^2}},
\end{equation}
where $\mu$ is the WiggleZ value of $D_2$ at radius $r$ and $\sigma$ is the root mean square variance given by the lognormal realisations.  See Fig. \ref{nr_pdfs_int1_D2_z57} for an illustration.

We also expect there to be be a likelihood distribution on the $D_2(r)$ value we would measure for a perfectly homogeneous distribution, due to cosmic variance and shot noise caused by our selection function. We can represent this by a likelihood distribution on the homogeneity scale -- this would not be a simple delta-function at $D_2(r) = 3$, but would have some spread. This distribution should be one-sided -- that is, we don't expect to measure $D_2(r) > 3$, only $D_2(r) \le 3$, if we have a distribution that approaches homogeneity.\footnote{In some cases the WiggleZ data does fall below $\mathcal{N} = 1$ or above $D_2 = 3$; this can be explained as the effect of shot noise introduced by the selection function rather than a physical effect, as shown in our comparison with the GiggleZ simulation in Section \ref{gigglez}.} We might expect it to be represented by the variance in the random catalogues, which are essentially homogeneous distributions sampled with the WiggleZ selection function. However, the same sources of noise are also present in the lognormal realisations, so we would potentially double-count errors if we also used lognormal realisations to determine the likelihood distribution of the data. This means we cannot easily determine the true variance in the value of $D_2(r)$ we would expect to measure for a homogeneous distribution, independently of the variance of the data.
 
 For this reason, we choose to find the likelihood distribution of the data reaching 1 per cent of homogeneity. That is, we assume the likelihood distribution on the homogeneity scale, $p_H[D_2(r)]$, is a delta-function at $D_2 = 2.97$:
\begin{equation}
{p_H[D_2(r)] \equiv \delta [D_2(r) -2.97]}.
\end{equation}

We can then construct the cumulative probability distribution function $P(R_H \le r)$, which gives the probability that the homogeneity scale has been reached at or before scale $r$, from:

\begin{eqnarray}
P(R_H \le r) &=& \int_{-\infty}^\infty p_D[D_2(r)] \left( \int^{D_2(r)}_{-\infty} p_H(x)dx\right) \textrm{d}D_2(r) \nonumber \\ 
&=& \int^{\infty}_{2.97} p_D[D_2(r)]\textrm{d}D_2(r).
\end{eqnarray}
That is, the probability of having reached homogeneity is the area of the likelihood distribution of the data that falls at or above $D_2=2.97$.

This cumulative probability is calculated at each scale $r$. We can then find the PDF for the homogeneity scale, $p(R_H)$, from
\begin{equation}
p(R_H) = {\textrm{d}P(R_H \le r) \over \textrm{d}r }.
\end{equation}

\begin{figure}
\includegraphics[width=9cm]{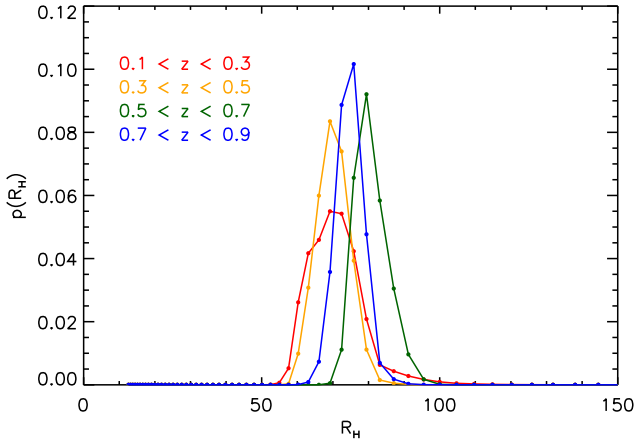}
\caption{Probability distributions for the scale of homogeneity, $p(R_H)$, for WiggleZ galaxies in each of the four redshift slices. The homogeneity scale is defined as the scale where the data reaches values 1 per cent away from $D_2=3$, i.e. $D_2=2.97$.   }
\label{nr_homscale_pdf_1win_int2_D2}
\end{figure}

The PDFs for the homogeneity scale, for WiggleZ galaxies in each redshift slice, are shown in Fig. \ref{nr_homscale_pdf_1win_int2_D2}. We have interpolated between the data bins in order to obtain smoother PDFs. We find the most probable $R_H$ values from the mean of the distributions. These are all between 70 and 81 $h^{-1}$ Mpc, and are listed in Table \ref{resultstable}. They represent the most probable scale at which the galaxy distribution reaches 1 per cent of homogeneity. We also list the values found for the bias-corrected data, which give the most probable homogeneity scales for the matter distribution, assuming $\Lambda$CDM+WMAP.

\section{Robustness of homogeneity measurement}
\label{robustness}

In this section we address several issues that could potentially influence our measurement of the homogeneity scale, and perform tests to ensure the robustness of our results. We base our tests on the 15-hr $0.5<z<0.7$ region, which is the largest and most populated WiggleZ sub-region, but the results are applicable to the entire survey.

\subsection{Fractal model test of selection function and boundary effects}
\label{fractal_test}

A major potential source of bias in our results is the method used to correct for edge effects and the selection function of the survey. We have used 100 random catalogues to correct each individual WiggleZ measurement, as described in Section \ref{nr}. However, this method can potentially bias homogeneity measurements, since it weights measurements by the volume of spheres of radius $r$ included in the survey, and so assumes a homogeneous distribution outside the survey \citep[e.g.][]{coleman1992,syloslabini2009}. It is therefore important to check that this is not imposing any distortion in our measured correlation dimension, so producing a `false relaxation' to homogeneity.

To test this we apply our correction method to a range of fractal distributions of known correlation dimension. This has been done previously by a number of works, e.g. \cite{lemson1991}, \cite{provenzale1994} and \cite{pan2002}. This allows us to check that our method returns the correct input correlation dimension up to the largest scales we measure in the survey, and to quantify any distortion that may occur. 

We generate our fractal distributions using the $\beta$-model, a simple self-similar cascading model \citep[see e.g.][]{castagnoli1991}. This method starts with a cube of side $L_0$ and splits it up into $M$ smaller cubes of side $L_0/n$ (we take $n=2$, so $M=8$). Each subcube is then assigned a probability $p$ of surviving to the next iteration. This is repeated for a certain number of iterations $k$, and the resultant set of survived points is taken as the final distribution. In the limit of an infinite number of iterations, this produces a monofractal with a correlation dimension given by
\begin{equation}
D_2= \lim_{k\to\infty} {\log(pM)^k \over \log n^k} = {\log pM \over \log n}.
\end{equation}

Our procedure for using $\beta$-models to test our analysis method is as follows:

\begin{enumerate}
	\item We choose a range of $D_2$ values (2.7, 2.8, 2.9, 2.95 and 2.97), and for each we generate 100 fractal galaxy distributions, with a boxsize of ($L_0$ $h^{-1}$Gpc)$^3$, where $L_0$ is the length of the longest side of the WiggleZ 15-hr $0.5<z<0.7$ selection function grid ({623.5 $h^{-1}$Mpc}). 
	\item We then sample each distribution with the WiggleZ selection function for the 15-hr $0.5<z<0.7$ region. We normalise the resulting distribution to give the same number of points as WiggleZ galaxies in this region. This gives us fractal mock catalogues, and we then measure $D_2$ for these in the same way as for the WiggleZ data, correcting the counts-in-spheres measurements with random catalogues. This gives a result that is influenced by both the WiggleZ selection function, and our correction method.
\end{enumerate}

\begin{figure*}
\includegraphics[width=18cm]{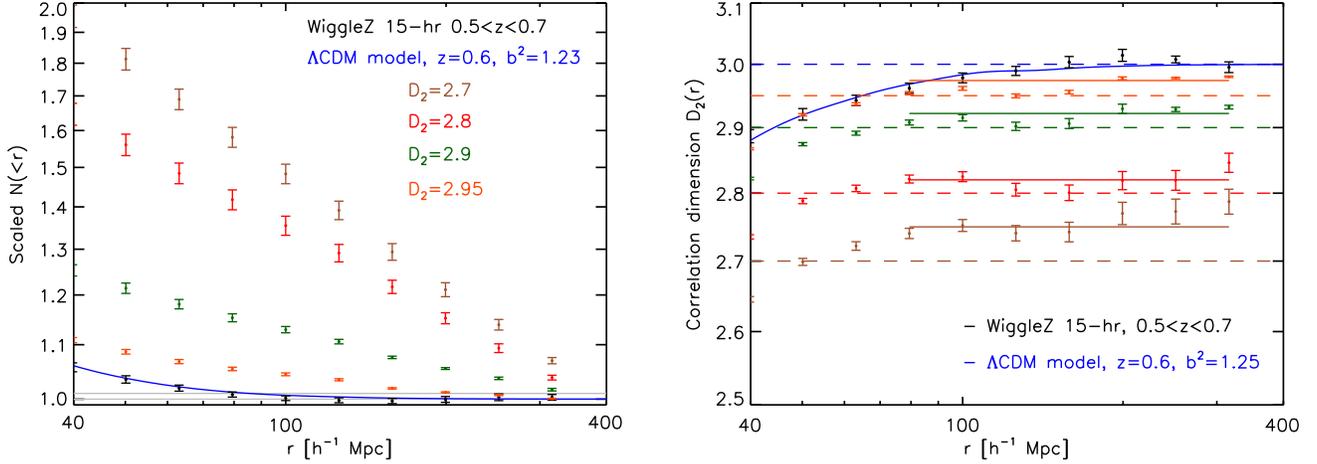}
\caption{Fractal model comparisons of the measured correlation dimension from the WiggleZ 15-hr region, $0.5<z<0.7$ redshift slice. The WiggleZ data (black error bars) and $\Lambda$CDM model (blue curve) are compared with several different $\beta$-models with different fractal dimension ($D_2=2.7,2.8,2.9$ and 2.95), which have been sampled with the 15-hr selection function and analysed in the same way as the WiggleZ data (coloured error bars). The uncertainties shown are the error-in-the-mean of 100 fractal realisations. The input fractal dimensions are shown as dotted lines with corresponding colours. The best-fit $D_2(r)$ values fit over the range $r=[80,300] h^{-1}$Mpc are shown as solid lines. }
\label{beta_nr_D2}
\end{figure*}

Fig. \ref{beta_nr_D2} shows the mean ${\mathcal{N}(<r)}$ and $D_2(r)$ results for the different fractal distributions up to $D_2=2.95$, with the WiggleZ selection function and correction method applied. Even up to $D_2=2.95$ they are clearly distinguishable from $\Lambda$CDM and the WiggleZ data.  The $D_2$ values measured are consistent with the input values up to at least 200 $h^{-1}$Mpc, well above the homogeneity scales measured for the data. These results indicate that the WiggleZ selection function and correction method do not have a significant effect on the measured correlation dimension up to the scales we measure for homogeneity. 

It is noticeable that the size of the error bars gets smaller for models with larger $D_2$. This is a real effect in the model; for larger $D_2$, more boxes survive at each iteration, so there are a smaller number of possible configurations for the final distribution, resulting in lower variance for a box of a particular volume. 

To quantify how well we can exclude fractal models, we fit a line of constant $D_2$ to each set of fractal data, over the range $[80,300]h^{-1}$Mpc (shown in Fig. \ref{beta_nr_D2}). This gives us the best-fit $D_2$ value we would expect to measure for each fractal distribution over this range, taking into account bias from the selection function. We then find the formal probability of these values fitting the WiggleZ data. Doing this, we find we can exclude a fractal dimension of $D_2(r)=[2.9,2.95,2.97]$ at the [19,6,4]-$\sigma$ level. In other words, we can exclude fractal distributions with dimension $D_2(r)<2.97$ at over 99.99 per cent confidence on scales from 80 to 300$h^{-1}$Mpc.

Our results agree with those of \cite{lemson1991}, \cite{provenzale1994} and \cite{pan2002} who also find that for samples on scales larger than the homogeneity scale, boundary corrections do not have a significant effect on the analysis. A further check would be to test different types of fractal model other than the $\beta$-model, but we leave this for future work. We also note that we still assume an FRW metric in our fractal analysis; an improved consistency check would be to calculate the actual metric in these fractal models (see Section \ref{discussion}), but this is beyond the scope of this paper.

\subsection{Comparison with the GiggleZ $N$-body simulation}
\label{gigglez}
We have also compared our results with a $\Lambda$CDM cosmological $N$-body simulation. This allows us to check that our analytical $\Lambda$CDM+WMAP model (incorporating Kaiser and streaming models for redshift-space distortions) is consistent, over the relevant scales, with a full simulation including nonlinear effects. It also provides a further test of selection function effects, since we can show that homogeneity measurements of a $\Lambda$CDM distribution are not distorted when the WiggleZ selection function is applied.

The GiggleZ (Giga-parsec WiggleZ) simulation (Poole et al. 2012, in preparation) is a suite of dark matter $N$-body simulations run at Swinburne University of Technology, designed for theoretical analyses of the WiggleZ dataset. It was run using a modified version of the {$N$-body} code \textsc{Gadget-2} \citep{springel2001} using a WMAP-5 cosmology.  We use the main simulation, which has a volume of ($1000 h^{-1}$Mpc)$^3$ and $2160^3$ particles of mass ${7.5 \times 10^9 h^{-1}M_\odot}$. 

Halo finding for GiggleZ was performed using Subfind \citep{springel2001}, which utilises a friends-of-friends (FoF) algorithm to identify coherent overdensities of particles and a substructure analysis to determine bound overdensities within each FoF halo. For our analysis, we rank-order the resulting Subfind substructure catalogues by their maximum circular velocity (V$_{\rm max}$) and select a contiguous subset of 250\,000 halos (selected to yield a number density comparable to the survey) over a range of V$_{\rm max}$ chosen to yield a bias comparable to that of WiggleZ. We use a catalogue at a redshift of ${z=0.593}$, corresponding to the mid redshift of the WiggleZ $0.5<z<0.7$ redshift slice.

We add the effect of redshift-space distortions by shifting the positions of the haloes according to their line-of-sight peculiar velocities. That is, the comoving position of each halo relative to an observer, $\textbf{x}$, is shifted by a vector $\Delta \textbf{x}$,
\begin{equation}
\Delta \textbf{x} = {\textbf{x} \over |\textbf{x}|}  {(1+z) \over H(z)} v_{\rm rad},
\end{equation}
where $v_{\rm rad} = (\textbf{x} \cdot \textbf{v}) / |\textbf{x}|$ is the radial velocity of the halo along the observer's line-of-sight, and we place the observer at the same coordinates relative to the GiggleZ box as for the WiggleZ selection function grid. 

We then calculate ${\mathcal{N}(<r)}$ and $D_2(r)$ using two different methods:
\begin{enumerate}

\item Using the full GiggleZ box. We correct the measurement using a random distribution within the same volume box, with 100 times the number of galaxies as the GiggleZ sample.

\item Applying the WiggleZ 15-hr ${0.5<z<0.7}$ selection function to GiggleZ. This creates a mock WiggleZ survey containing 10\,830 galaxies. We correct the measurement using the random catalogues used for the WiggleZ data. This allows us to see the effects induced purely by the selection function.

\end{enumerate}

\begin{figure*}
\includegraphics[width=18cm]{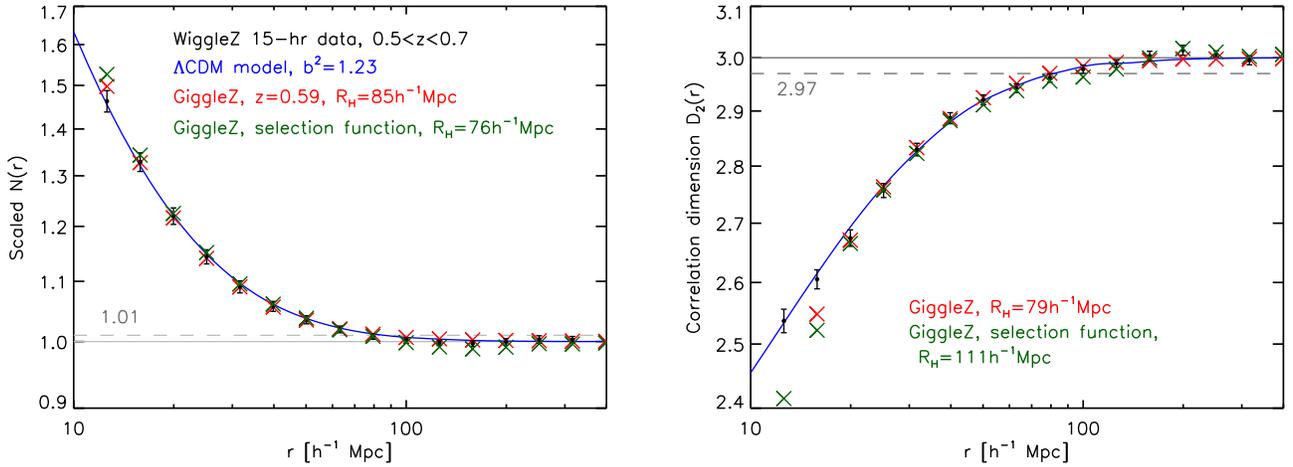}
\caption{Comparison of the GiggleZ $N$-body simulation with WiggleZ, for the 15-hr region ${0.5<z<0.7}$ redshift slice. The ${\mathcal{N}(<r)}$ results are shown on the left, and $D_2(r)$ on the right. The WiggleZ data is shown as black data points, and a $\Lambda$CDM model is shown in blue. The results for the full GiggleZ box are shown as the red crosses. The green crosses show the results for the GiggleZ simulation sampled with the WiggleZ 15-hr ${0.5<z<0.7}$ selection function. The measured homogeneity scale $R_H$ is indicated for each. }
\label{nrD2_gigglez2}
\end{figure*}

The results are shown in Fig. \ref{nrD2_gigglez2}. The full GiggleZ dataset is very consistent with both the WiggleZ data and the $\Lambda$CDM model. The consistency with the model indicates that the implementation of redshift-space distortions in the model, described in Section \ref{RSDs}, has a good level of accuracy, to scales as small as $\sim20h^{-1}$Mpc. The deviation of the smallest-scale bin is a resolution effect, since the GiggleZ catalogue is sparser than WiggleZ (with 10\,830 galaxies, vs. 17\,928 WiggleZ galaxies in the same region and redshift slice). The GiggleZ results both with, and without, the WiggleZ selection function are also consistent within the size of the WiggleZ error bars, showing that the selection function and correction method do not have a significant effect. It can be seen that the selection function causes a few data points in the ${\mathcal{N}(<r)}$ plot to go below 1, and in the $D_2(r)$ plot to go above 3. This shows that adding shot noise can produce this apparent unphysical effect, explaining why this is also seen in some of the WiggleZ results (Figs. \ref{nr_fit_101title} and \ref{D2_fit_299title}). 

\subsection{Comparison of different correction methods}
\label{corrections}

\begin{figure*}
\includegraphics[width=18cm]{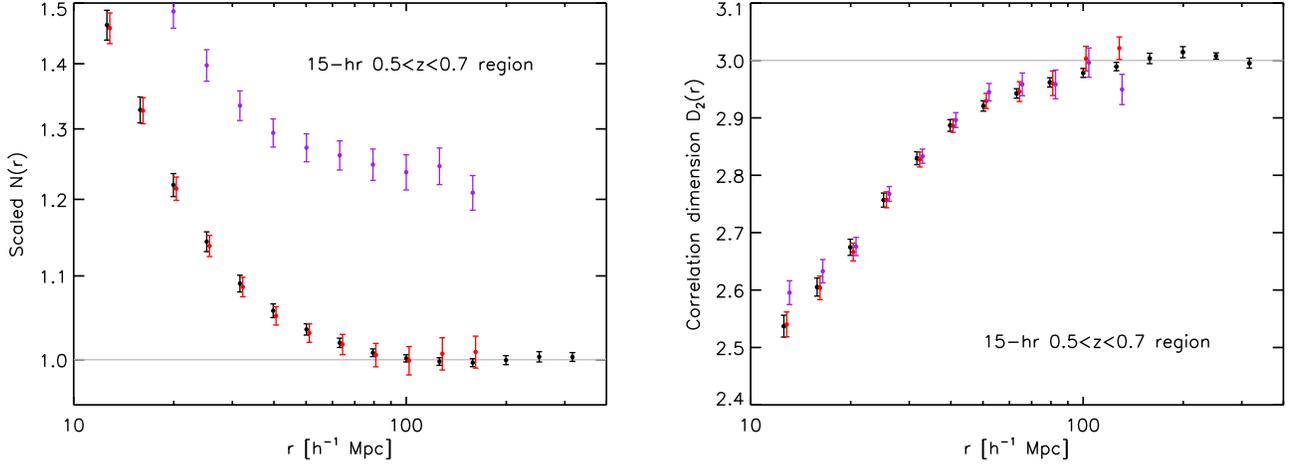}
\caption{Illustration of the result using different correction methods, for the 15-hr $0.5<z<0.7$ region. Black error bars show the result using our correction method with random catalogues. The red error bars are obtained when only complete spheres are used in the counts-in-spheres measurement, while still correcting for the selection function. The purple error bars show the result for complete spheres, with no selection function correction. In each case the uncertainties are calculated using lognormal realisations. The red and purple error bars are shifted slightly to the right for clarity.}
\label{Complete_spheres}
\end{figure*}

 To further demonstrate the robustness of our correction method, we illustrate the results obtained for two alternative correction methods -- firstly, using only complete spheres for the counts-in-spheres measurements (the `exclusion' method mentioned in Section \ref{nr}), but still correcting for the selection function, and secondly, using no correction for the selection function. By using the exclusion method we do not have to deal with survey edge effects;  however, since WiggleZ is not volume-limited, it is still necessary to use random catalogues to correct for the selection function. That is, we recalculate Eq. \ref{nr_eq} but with $G$ equal to the number of WiggleZ galaxies at the centre of complete spheres of radius $r$. We show this result in Fig.\ref{Complete_spheres}, for the 15-hr $0.5<z<0.7$ region, as red error bars, where the uncertainty is calculated using our lognormal realisations. We compare this to the result using our correction method (black error bars). The red error bars are consistent with the black error bars, but show more scatter and have higher noise, which increases for larger radius, reflecting the lower statistics where fewer spheres contribute to the measurement. The measurements must also be cut off at a lower radius, since not enough larger spheres fit inside the survey region.

We can also compare the result without any selection function correction, which illustrates the necessity of correcting for having a non-volume-limited sample. However, WiggleZ contains holes in the angular coverage, which are independent of assumptions about completion, so we must still take these into account. We therefore normalise each $N^i(r)$ measurement by the volume within the selection function included in that sphere. So we calculate:

\begin{equation}
\mathcal{N}_{\rm no\_corr}(r) = {1 \over G_{\rm complete}(r)}  \sum_{{\rm{complete\ spheres\ }}i }  { N^i(r) \over   V_{{\rm SF},i} \times \bar{\rho}  },
\end{equation}
where $G_{\rm complete}(r)$ is the number of WiggleZ galaxies at the centre of complete spheres of radius $r$, $V_{{\rm SF},i}$ is the volume within the selection function of the $i$th galaxy, and $\bar{\rho} = n_W / V_{\rm SF,tot}$ is the mean density of WiggleZ, i.e. the number of WiggleZ galaxies divided by the total volume in the selection function, excluding holes. We show this in Fig.\ref{Complete_spheres} as purple error bars, where the uncertainty is again calculated using our lognormal realisations (so the uncertainty assumes $\Lambda$CDM). Although there is a vertical offset in the ${\mathcal{N}(<r)}$ plot, caused by the selection function, the $D_2(r)$ results are remarkably similar to those when correcting for the selection function, though again with more noise. There is still a clear transition towards values close to $D_2=3$ on large scales. The offset in $\mathcal{N}(<r)$ can be attributed to a selection effect: since we use only complete spheres we weight the measurement towards the central part of the redshift range, where the completeness is highest and so the number density of WiggleZ galaxies is highest. Since $D_2(r)$ is the slope of $\mathcal{N}(<r)$, it does not depend on the number density itself and so is more robust when summing over a varying selection function.

\section{Discussion}
\label{discussion}

Our WiggleZ ${\mathcal{N}(<r)}$ and $D_2(r)$ results show a very strong agreement with an FRW-based $\Lambda$CDM model. However, one of the strongest arguments against previous homogeneity measurements is that the method of correcting for survey selection functions, such as using random catalogues as we do, can distort the data, producing a `false relaxation' to homogeneity. We have therefore tested this by applying the WiggleZ selection function to a range of inhomogeneous fractal models and a $\Lambda$CDM $N$-body simulation. We have shown that there is no significant impact on our homogeneity measurement, up to at least 200 $h^{-1}$Mpc. In addition, we have compared the results from our correction method to an analysis using only complete spheres, with and without correcting for the selection function, and have shown that these are consistent, further demonstrating the robustness of our result. We can rule out fractals with fractal dimension up to $D_2=2.97$, on scales between 80 and 300 $h^{-1}$Mpc, at over 99.99 per cent confidence.

We can also be confident that our result is robust to any assumptions in modelling the WiggleZ survey selection function, since \cite{blake2010} showed that even extreme variations in modelling the angular completeness produce only $\sim0.5\sigma$ shifts in estimates of the power spectrum. Changes in the parameterisation of the redshift distribution were shown to cause larger deviations, but only at scales $> 200\ h^{-1}$ Mpc, which are well above the scales on which we measure homogeneity.

Our result is a very good consistency check of $\Lambda$CDM. However, it is not independent of the assumption of the FRW metric. A complication for all homogeneity measurements is that we can only observe galaxies on our light cone, and not on spatial surfaces. \cite{maartens2011} points out that it is therefore not possible to make a homogeneity measurement without making some assumptions, such as the FRW metric (to convert redshifts to distances) and the Cosmological Principle. Indeed, we must always assume an FRW-based model, $\Lambda$CDM in our case, to convert redshifts and angles to distance coordinates. This is also problematic considering that inhomogeneities produce perturbations in the FRW metric, so can potentially distort distance measurements by affecting the paths travelled by light rays \citep[e.g.][]{wiltshire2009,meures2012}. 
However, it seems highly unlikely that our measurements would so closely agree with FRW-based $\Lambda$CDM (in both the amplitude and shape of the ${\mathcal{N}(<r)}$ and $D_2(r)$ curves), if the distribution were, actually, inhomogeneous up to the largest scales probed, and we had incorrectly assumed an FRW metric. We have tested our method using a fractal model, and shown that this gives a completely different form of these curves. Our assumption of FRW also seems reasonable, since we know that distance measurements from Type Ia supernovae fit the Hubble diagram well to first order up to $z\sim1.4$ \citep{conley2011,davis2011}, so that any perturbations due to inhomogeneities can only be a second- or higher-order effect. All this means that we can take our results to be a strong consistency check of FRW-based $\Lambda$CDM.

It would be possible to test this further by making isotropy measurements in thin redshift shells, over a range of redshifts, without converting them to distances. We could also directly test the assumption of FRW, by calculating the non-FRW metrics of our fractal models and testing our analysis on these. We could also calculate the effects of metric perturbations for $\Lambda$CDM, including the effects of backreaction, as in e.g.  \cite{buchert2000}, \cite{li2007}, \cite{wiltshire2007} and \cite{behrend2008}. This would allow us to test the effect of incorrectly assuming an FRW metric. 
Alternatively, several possible consistency tests of homogeneity and the Copernican Principle that do not assume FRW have been suggested by e.g. \cite{clarkson2008,shafieloo2010,maartens2011,heavens2011}. However, we leave these suggestions for future work.

An alternative way of defining the `homogeneity scale' was recently suggested by \cite{bagla2008} and \cite{yadav2010}, as where the measured fractal dimension $D_q$ becomes consistent with the ambient dimension within the 1$\sigma$ statistical uncertainty, $\sigma_{\Delta D_q}$. They make predictions for a $\Lambda$CDM model, by deriving an approximation for $D_q$ and $\sigma_{\Delta D_q}$, given a particular correlation function, and showing how these scale with sphere size. Using this, they predict that the `true' homogeneity scale in $\Lambda$CDM is 260 $h^{-1}$ Mpc. They also predict that $D_q$ and $\sigma_{\Delta D_q}$ scale the same way with the correlation function, and so their definition of homogeneity does not change with bias or redshift. Their definition is therefore beneficial, as it is robust to the tracer galaxy population. It is also not arbitrary, and indeed the scale above which the fractal dimension is consistent with homogeneity within cosmic variance is arguably a physically meaningful scale to define as homogeneous, since above this scale the distribution cannot be distinguished from a homogeneous one. However, as we have pointed out (Section \ref{rh}), it is difficult to apply their definition to a real measurement, since their approximation for $\sigma_{\Delta D_q}$ only accounts for the variance of the correlation function (in the limit of weak clustering) and shot noise, but ignores errors due to survey geometry and the selection function. These additional contributions mean that real errors will always be larger, and so will always measure a smaller homogeneity scale than the `true' one. Since these errors will be different for different surveys, and cannot be separated out from the variance in the underlying correlation function, homogeneity measurements made in this way cannot be easily compared between different surveys, or with theory. The benefit of our method of defining the homogeneity scale, even if we have to make an arbitrary choice about the value of $\Delta D_q$ we accept for homogeneity, is that it can be used to easily compare the results from different surveys and with a theoretical model.

A homogeneity scale below 100 $h^{-1}$Mpc may also seem to contradict the fact that the correlation function has a known feature, the Baryon Acoustic Oscillation (BAO) peak at $\sim105$ $h^{-1}$Mpc \citep{eisenstein2005,percival2010,beutler2011,blake2011c}. However, the BAO peak has only a small impact on the counts-in-spheres statistic. It is visible in our $\Lambda$CDM prediction for $D_2(r)$ (which is more sensitive to it than $\mathcal{N}(<r)$, since it is a differential measurement), as a small dip at just over 100 $h^{-1}$Mpc  (see Fig.\ref{analytic_nr_D2}). It means that the $D_2(r)$ curve does not increase monotonically around this scale, so we must be careful if attempting to measure an intercept with homogeneity that lies close to this. The magnitude of the distortion due to the BAO peak, $\Delta D_2$, is of order $\sim0.01$ for $b\sim1$ so does not affect our homogeneity scale measured at $D_2=2.97$. However, it is more significant for more highly-biased tracers \citep[as also pointed out by][]{bagla2008}, and for a highly-biased population, such as LRGs, it may be necessary to measure the homogeneity scale at a $D_2$ value closer to 3, to avoid this region. 

It has been pointed out \citep[e.g.][]{syloslabini2010} that measurements of large structures in galaxy surveys are seemingly at odds with a homogeneity scale below 100 $h^{-1}$Mpc. Previous surveys have detected structures on scales much larger than this \citep{geller1989,delapparent1989}. The largest observed structure in the Universe, the Sloan Great Wall, is 320 $h^{-1}$ Mpc long \citep{gott2005}, appears inconsistent with the existence of a homogeneity scale below 100 $h^{-1}$Mpc. However, this is not incompatible with our results, since we show the scale where the data has $D_2 \ge 2.97$, and it is not impossible to have fluctuations on scales larger than this. Also, these large structures, including the Sloan Great Wall, are usually filamentary, whereas we have measured a volume statistic which averages over fluctuations.

\section{Conclusion}
\label{conclusion}

We have measured the large-scale transition to homogeneity in the distribution of galaxies, using the WiggleZ survey, in four redshift bins between $z=0.1$ and $z=0.9$. We measured the mean, scaled counts-in-spheres ${\mathcal{N}(<r)}$ and the correlation dimension, $D_2(r)$, and found these to be in excellent agreement with a $\Lambda$CDM model with WMAP parameters, including redshift space distortions. We also presented a new, model-independent method for determining the homogeneity scale $R_H$ from data. This involves fitting a polynomial curve to the data, and finding where this intercepts chosen values close to homogeneity. This is a more reliable method than finding where the data comes within $1\sigma$ of homogeneity, since it does not depend directly on the size of the error bars and is less susceptible to noise. It also allows a direct comparison between data and theory, and between different surveys of differing bias and redshift.

We summarise our results as follows:
\begin{itemize}
\item Our ${\mathcal{N}(<r)}$ and $D_2(r)$ results show a very strong agreement with a FRW-based $\Lambda$CDM+WMAP model incorporating redshift-space distortions. They show a clear transition from an inhomogeneous, clustered distribution on small scales, to a homogeneous one on large scales. This transition matches that of the $\Lambda$CDM model. We have thereby conducted a very stringent consistency check of $\Lambda$CDM.
\item If we define the `homogeneity scale' $R_H$ as the scale where the data become consistent with homogeneity within 1 per cent, then from a likelihood analysis of $D_2(r)$, we measure $R_H$ to be $71 \pm8$ $h^{-1}$Mpc at $z\sim0.2$, $70 \pm 5$ $h^{-1}$Mpc at $z\sim0.4$, $81 \pm5$ $h^{-1}$Mpc at $z\sim0.6$, and $75 \pm4$ $h^{-1}$Mpc at $z\sim0.8$. These values are consistent with those of the $\Lambda$CDM+WMAP model with best-fitting bias, of $R_H$ = 76, 78, 81 and 78 $h^{-1}$Mpc.
\item We find that the homogeneity scale of our $\Lambda$CDM+WMAP model increases with clustering amplitude $b(z)\sigma_8(z)$. For a population with fixed bias, we therefore predict the homogeneity scale to grow over time, since $\sigma_8(z)$ increases due to growth of structure. The bias of the WiggleZ galaxies increases with redshift, so the measured $R_H$ values do not change with redshift. If we correct our data for bias, assuming $\Lambda$CDM, then we measure a homogeneity scale for the matter distribution that increases over time, consistent with our  $\Lambda$CDM+WMAP model.
\item The WiggleZ results are in excellent agreement with those of the GiggleZ $N$-body simulation incorporating redshift-space distortions. It is also in excellent agreement with our analytic $\Lambda$CDM+WMAP model, showing that our model for redshift-space distortions is accurate down to scales as small as $20h^{-1}$Mpc.
\item We can exclude a fractal with fractal dimension up to $D_2=2.97$, on scales between $\sim80h^{-1}$Mpc, and the largest scales probed by our measurement, $\sim300h^{-1}$Mpc, at 99.99 per cent confidence.
\item By applying our analysis to the GiggleZ simulation, as well as a suite of fractal distributions of differing fractal dimension, we have shown that our result is not significantly distorted by the WiggleZ selection function and our method of correcting for it. We also show that we obtain consistent results even using different correction methods, i.e. using only complete spheres for the measurement, with and without correcting for incompleteness. This therefore confirms the reliability and robustness of our results.
\end{itemize}

\section{Acknowledgements}

M.I.S. would like to thank Florian Beutler for many helpful comments on the paper. We also thank Eyal Kazin, David Parkinson, Luigi Guzzo and Andy Taylor for helpful discussions and comments.  M.I.S. acknowledges support from a Jean Rogerson Scholarship and a UWA Top-up Scholarship from the University of Western Australia. G.B.P. acknowledges support from two Australian Research Council (ARC) Discovery Projects (DP0772084 and DP1093738). The Centre for All-sky Astrophysics (CAASTRO) is an Australian Research Council Centre of Excellence, funded by grant CE11E0090. 

GALEX (the Galaxy Evolution Explorer) is a NASA Small Explorer, launched in April 2003. We gratefully ac- knowledge NASA's support for construction, operation and science analysis for the GALEX mission, developed in co- operation with the Centre National d'Etudes Spatiales of France and the Korean Ministry of Science and Technology.

The WiggleZ survey would not have been possible without the dedicated work of the staff of the Australian Astronomical Observatory in the development and support of the AAOmega spectrograph, and the running of the AAT.

\bibliographystyle{hapj}

\newcommand{\nat}{Nat}
\newcommand{\mnras}{MNRAS}
\newcommand{\apj}{ApJ}
\newcommand{\apjl}{ApJL}
\newcommand{\apjs}{ApJS}
\newcommand{\aap}{A \& A}
\newcommand{\aj}{AJ}
\newcommand{\pasa}{PASA}
\newcommand{\prd}{Phys. Rev. D}
\newcommand{\physrep}{Phys. Rep.}
\newcommand{\na}{NewA}
\newcommand{\jcap}{JCAP}

\bibliography{/Users/Morag/Write-ups/bibliography}

\begin{thebibliography}{90}
\expandafter\ifx\csname natexlab\endcsname\relax\def\natexlab#1{#1}\fi

\bibitem[{{Afshordi}(2004)}]{afshordi2004}
{Afshordi}, N. 2004, \prd, 70, 083536, arXiv:astro-ph/0401166

\bibitem[{{Amendola} \& {Palladino}(1999)}]{amendola1999}
{Amendola}, L., \& {Palladino}, E. 1999, \apjl, 514, L1, arXiv:astro-ph/9901420

\bibitem[{{Bagla} {et~al.}(2008){Bagla}, {Yadav}, \& {Seshadri}}]{bagla2008}
{Bagla}, J.~S., {Yadav}, J., \& {Seshadri}, T.~R. 2008, \mnras, 390, 829,
  0712.2905

\bibitem[{{Baumann}(2009)}]{baumann2009}
{Baumann}, D. 2009, ArXiv e-prints, 0907.5424

\bibitem[{{Behrend} {et~al.}(2008){Behrend}, {Brown}, \&
  {Robbers}}]{behrend2008}
{Behrend}, J., {Brown}, I.~A., \& {Robbers}, G. 2008, \jcap, 1, 13, 0710.4964

\bibitem[{{Beutler} {et~al.}(2011){Beutler}, {Blake}, {Colless}, {Jones},
  {Staveley-Smith}, {Campbell}, {Parker}, {Saunders}, \&
  {Watson}}]{beutler2011}
{Beutler}, F. {et~al.} 2011, \mnras, 416, 3017, 1106.3366

\bibitem[{{Blake} {et~al.}(2011{\natexlab{a}}){Blake}, {Brough}, {Colless},
  {Contreras}, {Couch}, {Croom}, {Davis}, {Drinkwater}, {Forster}, {Gilbank},
  {Gladders}, {Glazebrook}, {Jelliffe}, {Jurek}, {Li}, {Madore}, {Martin},
  {Pimbblet}, {Poole}, {Pracy}, {Sharp}, {Wisnioski}, {Woods}, {Wyder}, \&
  {Yee}}]{blake2011a}
{Blake}, C. {et~al.} 2011{\natexlab{a}}, \mnras, 415, 2876, 1104.2948

\bibitem[{{Blake} {et~al.}(2010){Blake}, {Brough}, {Colless}, {Couch}, {Croom},
  {Davis}, {Drinkwater}, {Forster}, {Glazebrook}, {Jelliffe}, {Jurek}, {Li},
  {Madore}, {Martin}, {Pimbblet}, {Poole}, {Pracy}, {Sharp}, {Wisnioski},
  {Woods}, \& {Wyder}}]{blake2010}
------. 2010, \mnras, 406, 803, 1003.5721

\bibitem[{{Blake} {et~al.}(2011{\natexlab{b}}){Blake}, {Davis}, {Poole},
  {Parkinson}, {Brough}, {Colless}, {Contreras}, {Couch}, {Croom},
  {Drinkwater}, {Forster}, {Gilbank}, {Gladders}, {Glazebrook}, {Jelliffe},
  {Jurek}, {Li}, {Madore}, {Martin}, {Pimbblet}, {Pracy}, {Sharp}, {Wisnioski},
  {Woods}, {Wyder}, \& {Yee}}]{blake2011b}
------. 2011{\natexlab{b}}, \mnras, 415, 2892, 1105.2862

\bibitem[{{Blake} {et~al.}(2009){Blake}, {Jurek}, {Brough}, {Colless}, {Couch},
  {Croom}, {Davis}, {Drinkwater}, {Forbes}, {Glazebrook}, {Madore}, {Martin},
  {Pimbblet}, {Poole}, {Pracy}, {Sharp}, {Small}, \& {Woods}}]{blake2009}
------. 2009, \mnras, 395, 240, 0901.2587

\bibitem[{{Blake} {et~al.}(2011{\natexlab{c}}){Blake}, {Kazin}, {Beutler},
  {Davis}, {Parkinson}, {Brough}, {Colless}, {Contreras}, {Couch}, {Croom},
  {Croton}, {Drinkwater}, {Forster}, {Gilbank}, {Gladders}, {Glazebrook},
  {Jelliffe}, {Jurek}, {Li}, {Madore}, {Martin}, {Pimbblet}, {Poole}, {Pracy},
  {Sharp}, {Wisnioski}, {Woods}, {Wyder}, \& {Yee}}]{blake2011c}
------. 2011{\natexlab{c}}, \mnras, 418, 1707, 1108.2635

\bibitem[{{Blake} \& {Wall}(2002)}]{blake2002}
{Blake}, C., \& {Wall}, J. 2002, \nat, 416, 150, arXiv:astro-ph/0203385

\bibitem[{{Borgani}(1995)}]{borgani1995}
{Borgani}, S. 1995, \physrep, 251, 1, arXiv:astro-ph/9404054

\bibitem[{{Brouzakis} {et~al.}(2007){Brouzakis}, {Tetradis}, \&
  {Tzavara}}]{brouzakis2007}
{Brouzakis}, N., {Tetradis}, N., \& {Tzavara}, E. 2007, \jcap, 2, 13,
  arXiv:astro-ph/0612179

\bibitem[{{Buchert}(2000)}]{buchert2000}
{Buchert}, T. 2000, General Relativity and Gravitation, 32, 105,
  arXiv:gr-qc/9906015

\bibitem[{{Buchert}(2008)}]{buchert2008}
------. 2008, General Relativity and Gravitation, 40, 467, 0707.2153

\bibitem[{{Castagnoli} \& {Provenzale}(1991)}]{castagnoli1991}
{Castagnoli}, C., \& {Provenzale}, A. 1991, \aap, 246, 634

\bibitem[{{Clarkson} {et~al.}(2008){Clarkson}, {Bassett}, \&
  {Lu}}]{clarkson2008}
{Clarkson}, C., {Bassett}, B., \& {Lu}, T.~H.-C. 2008, Physical Review Letters,
  101, 011301, 0712.3457

\bibitem[{{Clifton} {et~al.}(2008){Clifton}, {Ferreira}, \&
  {Land}}]{clifton2008}
{Clifton}, T., {Ferreira}, P.~G., \& {Land}, K. 2008, Physical Review Letters,
  101, 131302, 0807.1443

\bibitem[{{Coleman} \& {Pietronero}(1992)}]{coleman1992}
{Coleman}, P.~H., \& {Pietronero}, L. 1992, \physrep, 213, 311

\bibitem[{{Coles}(1993)}]{coles1993}
{Coles}, P. 1993, \mnras, 262, 1065

\bibitem[{{Coles} \& {Jones}(1991)}]{coles1991}
{Coles}, P., \& {Jones}, B. 1991, \mnras, 248, 1

\bibitem[{{Conley} {et~al.}(2011){Conley}, {Guy}, {Sullivan}, {Regnault},
  {Astier}, {Balland}, {Basa}, {Carlberg}, {Fouchez}, {Hardin}, {Hook},
  {Howell}, {Pain}, {Palanque-Delabrouille}, {Perrett}, {Pritchet}, {Rich},
  {Ruhlmann-Kleider}, {Balam}, {Baumont}, {Ellis}, {Fabbro}, {Fakhouri},
  {Fourmanoit}, {Gonz{\'a}lez-Gait{\'a}n}, {Graham}, {Hudson}, {Hsiao},
  {Kronborg}, {Lidman}, {Mourao}, {Neill}, {Perlmutter}, {Ripoche}, {Suzuki},
  \& {Walker}}]{conley2011}
{Conley}, A. {et~al.} 2011, \apjs, 192, 1, 1104.1443

\bibitem[{{Cowie} {et~al.}(1996){Cowie}, {Songaila}, {Hu}, \&
  {Cohen}}]{cowie1996}
{Cowie}, L.~L., {Songaila}, A., {Hu}, E.~M., \& {Cohen}, J.~G. 1996, \aj, 112,
  839, arXiv:astro-ph/9606079

\bibitem[{{Davis} {et~al.}(2011){Davis}, {Hui}, {Frieman}, {Haugb{\o}lle},
  {Kessler}, {Sinclair}, {Sollerman}, {Bassett}, {Marriner}, {M{\"o}rtsell},
  {Nichol}, {Richmond}, {Sako}, {Schneider}, \& {Smith}}]{davis2011}
{Davis}, T.~M. {et~al.} 2011, \apj, 741, 67, 1012.2912

\bibitem[{{de Lapparent} {et~al.}(1989){de Lapparent}, {Geller}, \&
  {Huchra}}]{delapparent1989}
{de Lapparent}, V., {Geller}, M.~J., \& {Huchra}, J.~P. 1989, \apj, 343, 1

\bibitem[{{Drinkwater} {et~al.}(2010){Drinkwater}, {Jurek}, {Blake}, {Woods},
  {Pimbblet}, {Glazebrook}, {Sharp}, {Pracy}, {Brough}, {Colless}, {Couch},
  {Croom}, {Davis}, {Forbes}, {Forster}, {Gilbank}, {Gladders}, {Jelliffe},
  {Jones}, {Li}, {Madore}, {Martin}, {Poole}, {Small}, {Wisnioski}, {Wyder}, \&
  {Yee}}]{drinkwater2010}
{Drinkwater}, M.~J. {et~al.} 2010, \mnras, 401, 1429, 0911.4246

\bibitem[{{Eisenstein} {et~al.}(2005){Eisenstein}, {Zehavi}, {Hogg},
  {Scoccimarro}, {Blanton}, {Nichol}, {Scranton}, {Seo}, {Tegmark}, {Zheng},
  {Anderson}, {Annis}, {Bahcall}, {Brinkmann}, {Burles}, {Castander},
  {Connolly}, {Csabai}, {Doi}, {Fukugita}, {Frieman}, {Glazebrook}, {Gunn},
  {Hendry}, {Hennessy}, {Ivezi{\'c}}, {Kent}, {Knapp}, {Lin}, {Loh}, {Lupton},
  {Margon}, {McKay}, {Meiksin}, {Munn}, {Pope}, {Richmond}, {Schlegel},
  {Schneider}, {Shimasaku}, {Stoughton}, {Strauss}, {SubbaRao}, {Szalay},
  {Szapudi}, {Tucker}, {Yanny}, \& {York}}]{eisenstein2005}
{Eisenstein}, D.~J. {et~al.} 2005, \apj, 633, 560, arXiv:astro-ph/0501171

\bibitem[{{Ellis} \& {Buchert}(2005)}]{ellis2005}
{Ellis}, G.~F.~R., \& {Buchert}, T. 2005, Physics Letters A, 347, 38,
  arXiv:gr-qc/0506106

\bibitem[{{Fisher}(1995)}]{fisher1995}
{Fisher}, K.~B. 1995, \apj, 448, 494, arXiv:astro-ph/9412081

\bibitem[{{Fixsen} {et~al.}(1996){Fixsen}, {Cheng}, {Gales}, {Mather},
  {Shafer}, \& {Wright}}]{fixsen1996}
{Fixsen}, D.~J., {Cheng}, E.~S., {Gales}, J.~M., {Mather}, J.~C., {Shafer},
  R.~A., \& {Wright}, E.~L. 1996, \apj, 473, 576, arXiv:astro-ph/9605054

\bibitem[{{Gabrielli} {et~al.}(2002){Gabrielli}, {Joyce}, \& {Sylos
  Labini}}]{gabrielli2002}
{Gabrielli}, A., {Joyce}, M., \& {Sylos Labini}, F. 2002, \prd, 65, 083523,
  arXiv:astro-ph/0110451

\bibitem[{{Geller} \& {Huchra}(1989)}]{geller1989}
{Geller}, M.~J., \& {Huchra}, J.~P. 1989, Science, 246, 897

\bibitem[{{Glazebrook} {et~al.}(2004){Glazebrook}, {Abraham}, {McCarthy},
  {Savaglio}, {Chen}, {Crampton}, {Murowinski}, {J{\o}rgensen}, {Roth}, {Hook},
  {Marzke}, \& {Carlberg}}]{glazebrook2004}
{Glazebrook}, K. {et~al.} 2004, \nat, 430, 181, arXiv:astro-ph/0401037

\bibitem[{{Gott} {et~al.}(2005){Gott}, {Juri{\'c}}, {Schlegel}, {Hoyle},
  {Vogeley}, {Tegmark}, {Bahcall}, \& {Brinkmann}}]{gott2005}
{Gott}, III, J.~R., {Juri{\'c}}, M., {Schlegel}, D., {Hoyle}, F., {Vogeley},
  M., {Tegmark}, M., {Bahcall}, N., \& {Brinkmann}, J. 2005, \apj, 624, 463,
  arXiv:astro-ph/0310571

\bibitem[{{Guzzo}(1997)}]{guzzo1997}
{Guzzo}, L. 1997, \na, 2, 517, arXiv:astro-ph/9711206

\bibitem[{{Hamilton}(1992)}]{hamilton1992}
{Hamilton}, A.~J.~S. 1992, \apjl, 385, L5

\bibitem[{{Hamilton}(1993)}]{hamilton1993}
------. 1993, \apj, 417, 19

\bibitem[{{Hatton} \& {Cole}(1998)}]{hatton1998}
{Hatton}, S., \& {Cole}, S. 1998, \mnras, 296, 10, arXiv:astro-ph/9707186

\bibitem[{{Hawkins} {et~al.}(2003){Hawkins}, {Maddox}, {Cole}, {Lahav},
  {Madgwick}, {Norberg}, {Peacock}, {Baldry}, {Baugh}, {Bland-Hawthorn},
  {Bridges}, {Cannon}, {Colless}, {Collins}, {Couch}, {Dalton}, {De Propris},
  {Driver}, {Efstathiou}, {Ellis}, {Frenk}, {Glazebrook}, {Jackson}, {Jones},
  {Lewis}, {Lumsden}, {Percival}, {Peterson}, {Sutherland}, \&
  {Taylor}}]{hawkins2003}
{Hawkins}, E. {et~al.} 2003, \mnras, 346, 78, arXiv:astro-ph/0212375

\bibitem[{{Heavens} {et~al.}(2011){Heavens}, {Jimenez}, \&
  {Maartens}}]{heavens2011}
{Heavens}, A.~F., {Jimenez}, R., \& {Maartens}, R. 2011, \jcap, 9, 35,
  1107.5910

\bibitem[{{Hogg} {et~al.}(2005){Hogg}, {Eisenstein}, {Blanton}, {Bahcall},
  {Brinkmann}, {Gunn}, \& {Schneider}}]{hogg2005}
{Hogg}, D.~W., {Eisenstein}, D.~J., {Blanton}, M.~R., {Bahcall}, N.~A.,
  {Brinkmann}, J., {Gunn}, J.~E., \& {Schneider}, D.~P. 2005, \apj, 624, 54,
  arXiv:astro-ph/0411197

\bibitem[{{Joyce} {et~al.}(2000){Joyce}, {Anderson}, {Montuori}, {Pietronero},
  \& {Sylos Labini}}]{joyce2000}
{Joyce}, M., {Anderson}, P.~W., {Montuori}, M., {Pietronero}, L., \& {Sylos
  Labini}, F. 2000, Europhysics Letters, 50, 416, arXiv:astro-ph/0002504

\bibitem[{{Joyce} {et~al.}(1999){Joyce}, {Montuori}, \& {Sylos
  Labini}}]{joyce1999}
{Joyce}, M., {Montuori}, M., \& {Sylos Labini}, F. 1999, \apjl, 514, L5,
  arXiv:astro-ph/9901290

\bibitem[{{Kaiser}(1984)}]{kaiser1984}
{Kaiser}, N. 1984, \apjl, 284, L9

\bibitem[{{Kaiser}(1987)}]{kaiser1987}
------. 1987, \mnras, 227, 1

\bibitem[{{Kolb} {et~al.}(2005){Kolb}, {Matarrese}, {Notari}, \&
  {Riotto}}]{kolb2005}
{Kolb}, E.~W., {Matarrese}, S., {Notari}, A., \& {Riotto}, A. 2005, \prd, 71,
  023524, arXiv:hep-ph/0409038

\bibitem[{{Komatsu} {et~al.}(2011){Komatsu}, {Smith}, {Dunkley}, {Bennett},
  {Gold}, {Hinshaw}, {Jarosik}, {Larson}, {Nolta}, {Page}, {Spergel},
  {Halpern}, {Hill}, {Kogut}, {Limon}, {Meyer}, {Odegard}, {Tucker}, {Weiland},
  {Wollack}, \& {Wright}}]{komatsu2011}
{Komatsu}, E. {et~al.} 2011, \apjs, 192, 18, 1001.4538

\bibitem[{{Kurokawa} {et~al.}(2001){Kurokawa}, {Morikawa}, \&
  {Mouri}}]{kurokawa2001}
{Kurokawa}, T., {Morikawa}, M., \& {Mouri}, H. 2001, \aap, 370, 358

\bibitem[{{Landy} \& {Szalay}(1993)}]{landy1993}
{Landy}, S.~D., \& {Szalay}, A.~S. 1993, \apj, 412, 64

\bibitem[{{Lemson} \& {Sanders}(1991)}]{lemson1991}
{Lemson}, G., \& {Sanders}, R.~H. 1991, \mnras, 252, 319

\bibitem[{{Lewis} {et~al.}(2000){Lewis}, {Challinor}, \& {Lasenby}}]{lewis2000}
{Lewis}, A., {Challinor}, A., \& {Lasenby}, A. 2000, \apj, 538, 473,
  arXiv:astro-ph/9911177

\bibitem[{{Li} \& {Schwarz}(2007)}]{li2007}
{Li}, N., \& {Schwarz}, D.~J. 2007, \prd, 76, 083011, arXiv:gr-qc/0702043

\bibitem[{{Maartens}(2011)}]{maartens2011}
{Maartens}, R. 2011, Royal Society of London Philosophical Transactions Series
  A, 369, 5115, 1104.1300

\bibitem[{{Mart\'{i}nez} \& {Saar}(2002)}]{martinez2002}
{Mart\'{i}nez}, V., \& {Saar}, E. 2002, in Astronomical Data Analysis II.
  Edited by Starck, Jean-Luc; Murtagh, Fionn D. Proceedings of the SPIE, Volume
  4847, pp. 86-100 (2002)., ed. J.-L. {Starck} \& F.~D. {Murtagh}, Vol. 4847,
  86--100, arXiv:astro-ph/0209208

\bibitem[{{Mart\'{i}nez} \& {Coles}(1994)}]{martinez1994}
{Mart\'{i}nez}, V.~J., \& {Coles}, P. 1994, \apj, 437, 550

\bibitem[{{Mart\'{i}nez} {et~al.}(1998){Mart\'{i}nez}, {Pons-Border\'{i}a},
  {Moyeed}, \& {Graham}}]{martinez1998}
{Mart\'{i}nez}, V.~J., {Pons-Border\'{i}a}, M.-J., {Moyeed}, R.~A., \&
  {Graham}, M.~J. 1998, \mnras, 298, 1212, arXiv:astro-ph/9804073

\bibitem[{{Meures} \& {Bruni}(2012)}]{meures2012}
{Meures}, N., \& {Bruni}, M. 2012, \mnras, 419, 1937, 1107.4433

\bibitem[{{Pan} \& {Coles}(2000)}]{pan2000}
{Pan}, J., \& {Coles}, P. 2000, \mnras, 318, L51, arXiv:astro-ph/0008240

\bibitem[{{Pan} \& {Coles}(2002)}]{pan2002}
------. 2002, \mnras, 330, 719, arXiv:astro-ph/0111234

\bibitem[{{Peacock}(1999)}]{peacock1999}
{Peacock}, J.~A. 1999, {Cosmological Physics}, ed. {Peacock, J.~A.}

\bibitem[{{Peacock} \& {Dodds}(1994)}]{peacock1994}
{Peacock}, J.~A., \& {Dodds}, S.~J. 1994, \mnras, 267, 1020,
  arXiv:astro-ph/9311057

\bibitem[{{Peebles}(1980)}]{peebles1980}
{Peebles}, P.~J.~E. 1980, {The large-scale structure of the universe}

\bibitem[{{Peebles}(1993)}]{peebles1993}
------. 1993, {Principles of Physical Cosmology} (Princeton University Press)

\bibitem[{{Percival} {et~al.}(2010){Percival}, {Reid}, {Eisenstein}, {Bahcall},
  {Budavari}, {Frieman}, {Fukugita}, {Gunn}, {Ivezi{\'c}}, {Knapp}, {Kron},
  {Loveday}, {Lupton}, {McKay}, {Meiksin}, {Nichol}, {Pope}, {Schlegel},
  {Schneider}, {Spergel}, {Stoughton}, {Strauss}, {Szalay}, {Tegmark},
  {Vogeley}, {Weinberg}, {York}, \& {Zehavi}}]{percival2010}
{Percival}, W.~J. {et~al.} 2010, \mnras, 401, 2148, 0907.1660

\bibitem[{{Pietronero} {et~al.}(1997){Pietronero}, {Montuori}, \& {Sylos
  Labini}}]{pietronero1997}
{Pietronero}, L., {Montuori}, M., \& {Sylos Labini}, F. 1997, in Critical
  Dialogues in Cosmology, ed. {N.~Turok}, 24--27, arXiv:astro-ph/9611197

\bibitem[{{Provenzale} {et~al.}(1994){Provenzale}, {Guzzo}, \&
  {Murante}}]{provenzale1994}
{Provenzale}, A., {Guzzo}, L., \& {Murante}, G. 1994, \mnras, 266, 555

\bibitem[{{R{\"a}s{\"a}nen}(2006)}]{rasanen2006}
{R{\"a}s{\"a}nen}, S. 2006, \jcap, 11, 3, arXiv:astro-ph/0607626

\bibitem[{{R{\"a}s{\"a}nen}(2011)}]{rasanen2011}
------. 2011, Classical and Quantum Gravity, 28, 164008, 1102.0408

\bibitem[{Ripley(1977)}]{ripley1977}
Ripley, B.~D. 1977, Journal of the Royal Statistical Society. Series B
  (Methodological), 39, pp. 172

\bibitem[{{Sachs} \& {Wolfe}(1967)}]{sachs1967}
{Sachs}, R.~K., \& {Wolfe}, A.~M. 1967, \apj, 147, 73

\bibitem[{{Sarkar} {et~al.}(2009){Sarkar}, {Yadav}, {Pandey}, \&
  {Bharadwaj}}]{sarkar2009}
{Sarkar}, P., {Yadav}, J., {Pandey}, B., \& {Bharadwaj}, S. 2009, \mnras, 399,
  L128, 0906.3431

\bibitem[{{Scaramella} {et~al.}(1998){Scaramella}, {Guzzo}, {Zamorani},
  {Zucca}, {Balkowski}, {Blanchard}, {Cappi}, {Cayatte}, {Chincarini},
  {Collins}, {Fiorani}, {Maccagni}, {MacGillivray}, {Maurogordato}, {Merighi},
  {Mignoli}, {Proust}, {Ramella}, {Stirpe}, \& {Vettolani}}]{scaramella1998}
{Scaramella}, R. {et~al.} 1998, \aap, 334, 404, arXiv:astro-ph/9803022

\bibitem[{{Scharf} {et~al.}(2000){Scharf}, {Jahoda}, {Treyer}, {Lahav},
  {Boldt}, \& {Piran}}]{scharf2000}
{Scharf}, C.~A., {Jahoda}, K., {Treyer}, M., {Lahav}, O., {Boldt}, E., \&
  {Piran}, T. 2000, \apj, 544, 49, arXiv:astro-ph/9908187

\bibitem[{{Scherrer} \& {Weinberg}(1998)}]{scherrer1998}
{Scherrer}, R.~J., \& {Weinberg}, D.~H. 1998, \apj, 504, 607,
  arXiv:astro-ph/9712192

\bibitem[{{Schwarz}(2002)}]{schwarz2002}
{Schwarz}, D.~J. 2002, ArXiv Astrophysics e-prints, arXiv:astro-ph/0209584

\bibitem[{{Shafieloo} \& {Clarkson}(2010)}]{shafieloo2010}
{Shafieloo}, A., \& {Clarkson}, C. 2010, \prd, 81, 083537, 0911.4858

\bibitem[{{Smith} {et~al.}(2003){Smith}, {Peacock}, {Jenkins}, {White},
  {Frenk}, {Pearce}, {Thomas}, {Efstathiou}, \& {Couchman}}]{smith2003}
{Smith}, R.~E. {et~al.} 2003, \mnras, 341, 1311, arXiv:astro-ph/0207664

\bibitem[{{Springel} {et~al.}(2001){Springel}, {Yoshida}, \&
  {White}}]{springel2001}
{Springel}, V., {Yoshida}, N., \& {White}, S.~D.~M. 2001, \na, 6, 79,
  arXiv:astro-ph/0003162

\bibitem[{{Sylos Labini}(2011)}]{syloslabini2011}
{Sylos Labini}, F. 2011, EPL (Europhysics Letters), 96, 59001, 1110.4041

\bibitem[{{Sylos Labini} {et~al.}(1998){Sylos Labini}, {Montuori}, \&
  {Pietronero}}]{syloslabini1998}
{Sylos Labini}, F., {Montuori}, M., \& {Pietronero}, L. 1998, \physrep, 293,
  61, arXiv:astro-ph/9711073

\bibitem[{{Sylos Labini} \& {Pietronero}(2010)}]{syloslabini2010}
{Sylos Labini}, F., \& {Pietronero}, L. 2010, Journal of Statistical Mechanics:
  Theory and Experiment, 11, 29, 1012.5624

\bibitem[{{Sylos Labini} {et~al.}(2009){Sylos Labini}, {Vasilyev}, \&
  {Baryshev}}]{syloslabini2009}
{Sylos Labini}, F., {Vasilyev}, N.~L., \& {Baryshev}, Y.~V. 2009, \aap, 508,
  17, 0909.0132

\bibitem[{{Wiltshire}(2007{\natexlab{a}})}]{wiltshire2007a}
{Wiltshire}, D.~L. 2007{\natexlab{a}}, New Journal of Physics, 9, 377,
  arXiv:gr-qc/0702082

\bibitem[{{Wiltshire}(2007{\natexlab{b}})}]{wiltshire2007}
------. 2007{\natexlab{b}}, Physical Review Letters, 99, 251101, 0709.0732

\bibitem[{{Wiltshire}(2009)}]{wiltshire2009}
------. 2009, \prd, 80, 123512, 0909.0749

\bibitem[{{Wu} {et~al.}(1999){Wu}, {Lahav}, \& {Rees}}]{wu1999}
{Wu}, K.~K.~S., {Lahav}, O., \& {Rees}, M.~J. 1999, \nat, 397, 225,
  arXiv:astro-ph/9804062

\bibitem[{{Yadav} {et~al.}(2005){Yadav}, {Bharadwaj}, {Pandey}, \&
  {Seshadri}}]{yadav2005}
{Yadav}, J., {Bharadwaj}, S., {Pandey}, B., \& {Seshadri}, T.~R. 2005, \mnras,
  364, 601, arXiv:astro-ph/0504315

\bibitem[{{Yadav} {et~al.}(2010){Yadav}, {Bagla}, \& {Khandai}}]{yadav2010}
{Yadav}, J.~K., {Bagla}, J.~S., \& {Khandai}, N. 2010, \mnras, 405, 2009,
  1001.0617

\bibitem[{{Zehavi} {et~al.}(2005){Zehavi}, {Zheng}, {Weinberg}, {Frieman},
  {Berlind}, {Blanton}, {Scoccimarro}, {Sheth}, {Strauss}, {Kayo}, {Suto},
  {Fukugita}, {Nakamura}, {Bahcall}, {Brinkmann}, {Gunn}, {Hennessy},
  {Ivezi{\'c}}, {Knapp}, {Loveday}, {Meiksin}, {Schlegel}, {Schneider},
  {Szapudi}, {Tegmark}, {Vogeley}, {York}, \& {SDSS
  Collaboration}}]{zehavi2005}
{Zehavi}, I. {et~al.} 2005, \apj, 630, 1, arXiv:astro-ph/0408569

\end{thebibliography}

\appendix
\section{Gaussian vs true probability distributions $p_D[D_2(r)]$}
\label{pdfappendix}

In our likelihood analysis for the homogeneity scale (Section \ref{likelihood}) we assume that the distribution of the 100 lognormal realisations in each bin, $p_D[D_2(r)]$, is Gaussian. This allows us to interpolate between the data points and errors, so create more finely-spaced $p_D[D_2(r)]$ distributions, in order to determine a smoother PDF for the homogeneity scale, $P(R_H\le r )$. However, it is not obvious that these distributions should be Gaussian. We therefore repeat the analysis, but use the true distributions given by the lognormal realisations.

\begin{figure}
\includegraphics[width=9cm]{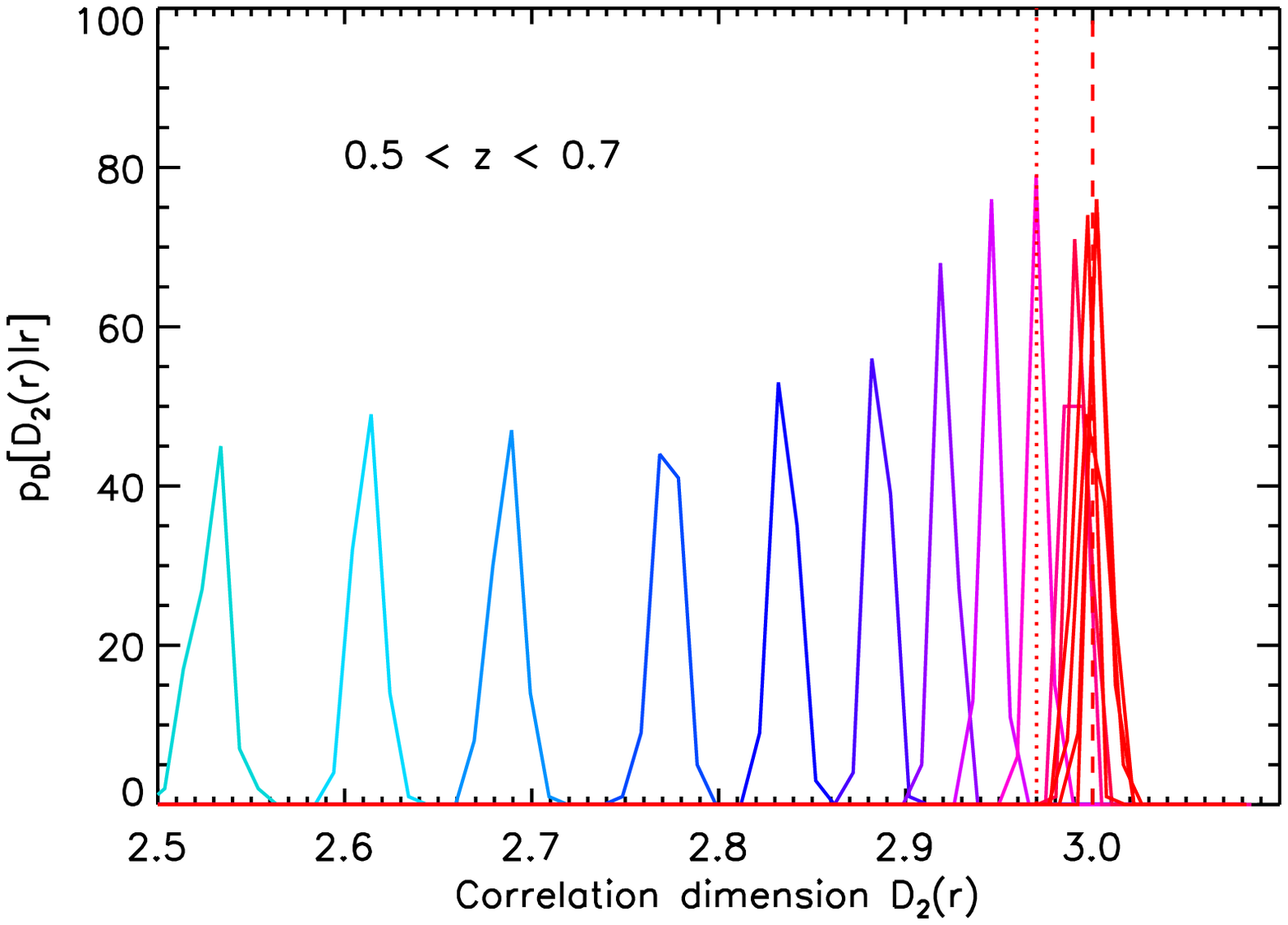}
\caption{The probability distributions $p_D[D_2(r)]$ for each of the $r$ bins (blue-to-red gradient indicates small to large radius) in the $0.5<z<0.7$ redshift slice.  Unlike in Fig.\ref{nr_pdfs_int1_D2_z57}, we do not assume these are Gaussian distributions; rather, we plot the distributions given by our 100 lognormal realisations.}
\label{D2_pdfs_lognormals3}
\end{figure}

This gives the $p_D[D_2(r)]$ distributions shown in Fig. \ref{D2_pdfs_lognormals3}. They are not smooth Gaussians, although they are close to Gaussian. Their resolution is limited by the number of lognormal realisations, so they could be improved by using more lognormal realisations, although we do not do this here.

\begin{figure}
\includegraphics[width=9cm]{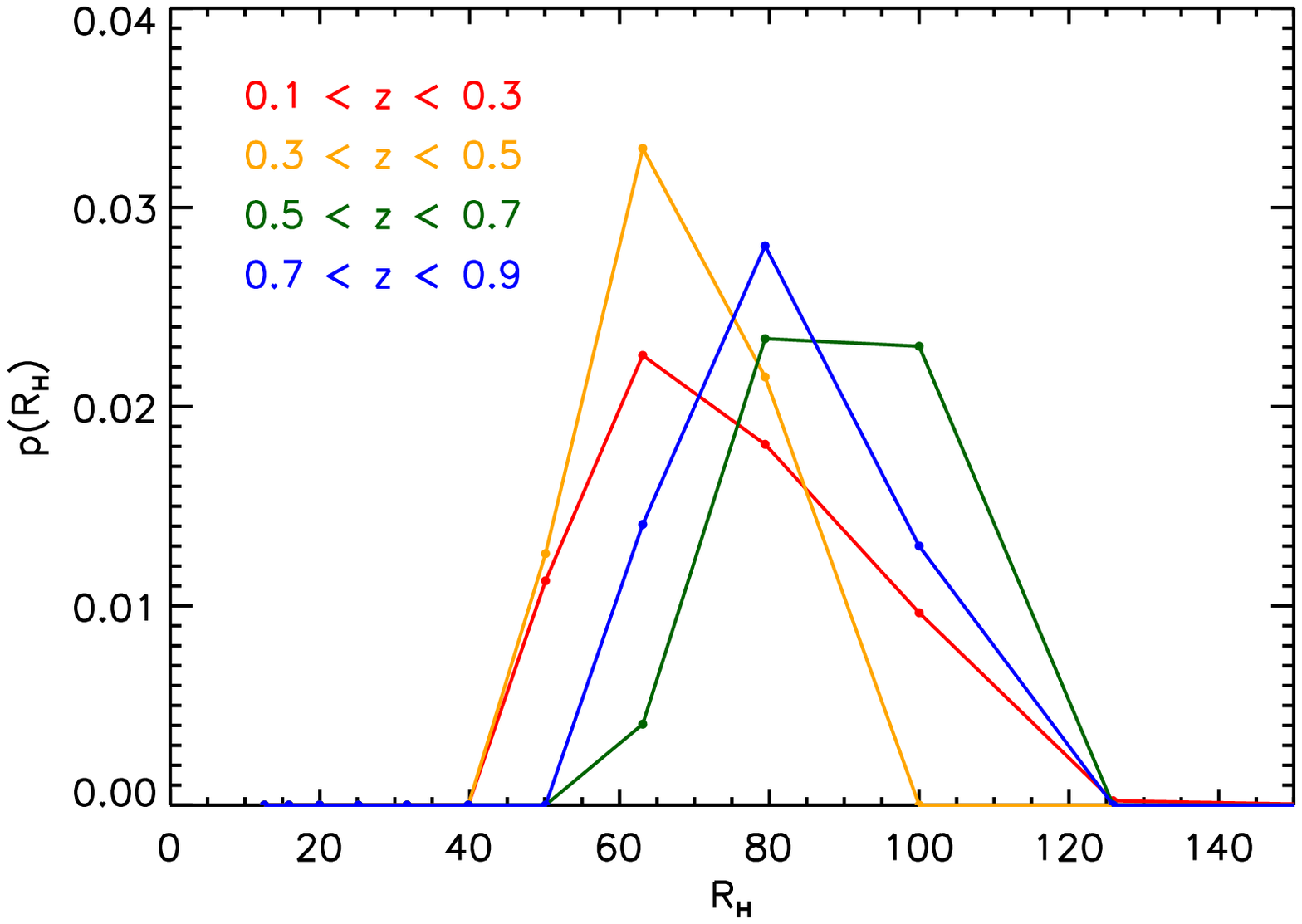}
\caption{Probability distributions for the homogeneity scale, $p(R_H)$, for WiggleZ galaxies in each of the four redshift slices. These are calculated from the probability distributions $p_D[D_2(r)]$ of the 100 lognormal realisations, rather than assuming Gaussians.}
\label{homscale_D2_lognormals1}
\end{figure}

We then use these to calculate the PDF for the homogeneity scale, $P(R_H\le r )$, in the same way as we did in Section \ref{likelihood} for the Gaussian distributions. This gives the PDFs shown in Fig.\ref{homscale_D2_lognormals1}, for each redshift slice. The mean values and errors are shown in Table \ref{pdfcomp}, along with those from our original analysis. The values are very similar, though the errors are larger. This is because we cannot interpolate between data points as easily, and there are a finite number of lognormal realisations contributing to the distribution for each datapoint. This means the distribution is effectively smoothed, giving larger uncertainties.

 \begin{table}
 \caption{Comparison of the most probable $R_H$ values from a likelihood analysis using the true $p_D[D_2(r)]$ distributions from lognormal realisations, and assuming Gaussian distributions.}
 \label{pdfcomp}
 \begin{tabular}{@{}lcccccc}
  \hline
Redshift & $R_H$ for true $p_D[D_2(r)]$  &  $R_H$ assuming Gaussians \\
& distributions [$h^{-1}$Mpc] & [$h^{-1}$Mpc] \\
  \hline
  $0.1<z<0.3$ & $79\pm19$ &  $71\pm8$\\
$0.3<z<0.5$ & $71\pm13$ & $70\pm5$\\
 $0.5<z<0.7$ & $91\pm16$& $81\pm5$\\
$0.7<z<0.9$ & $84\pm16$  &  $75\pm4$\\
  \hline
 \end{tabular}
 \end{table}


\end{document}